\documentclass[prb,showpacs,floatfix]{revtex4}
\usepackage{graphicx}
\usepackage{dcolumn}
\usepackage{bm}

\begin{document}

\preprint{}

\title{The long-term cyclotron dynamics of relativistic wave packets: spontaneous collapse and revival}

\author{V.Ya.Demikhovskii, G.M.Maksimova, A.A.Perov, and A.V.Telezhnikov}

\affiliation{Department of Physics, University of Nizhny Novgorod,
              23 Gagarin Avenue, 603950 Nizhny Novgorod, Russian Federation}

\date{\today}

\begin{abstract}

In this work we study the effects of collapse and revival as well
as {\it Zitterbewegung} (ZB) phenomenon, for the relativistic
electron wave packets, which are a superposition of the states
with quantum numbers sharply peaked around some Landau level $n_0$
of the order of few tens. The probability densities as well as
average velocities of the packet center and the average spin
components were calculated analytically and visualized. Our
computations demonstrate that due to dephasing of the states for
times larger than the cyclotron period the initial wave packet
(which includes the states with the positive energy only) loses
the spatial localization so that the evolution can no longer be
described classically. However, at the half-revival time $t=T_R/2$
its reshaping takes place firstly. The behavior of the wave packet
containing the states of both energy bands (with $E_n>0$ and
$E_n<0$) is more complicated. At short times of a few classical
periods such packet splits into two parts which rotate with
cyclotron frequency in the opposite directions and meet each other
every one-half of the cyclotron period. At these moments their
wave functions have significant overlap that leads to ZB. At the
time of fractional revival each of two sub-packets is decomposed
into few packet-fractions. However, at $t=T_R$ each of the two
sub-packets (with positive or negative energy) restores at various
points of the cyclotron orbit, that makes it impossible reshaping
of initial wave packet entirely unlike the wave packet which
consists of states with energies $E_n>0$ only. Obtained results
can be useful for the description of electromagnetic radiation and
absorption in relativistic plasma on astrophysics objects, where
super high magnetic field has the value of the order $10^8-10^9$T,
as well as for interpretation of experiments with trapped ions.

\end{abstract}


\maketitle

\section{Introduction}
The space-time evolution of quantum system with discrete but not equidistant energy spectrum can be quite complex.
It was shown\cite{A1} that for the initial state consisting of the eigenstates with quantum numbers $n$
sharply peaked around some level $n_0\gg 1$ the coefficients of Taylor expansion of spectrum
$E_n$ around that value define the different time scales of wave packet periodic evolution.

At the beginning a localized wave packet moves periodically along
a classical trajectory with the period
$T_{cl}=2\pi\hbar/E'_{n_0}$. However, over time, the nonlinear
terms in $(n-n_0)$ in the expansion of the spectrum $E_n$ become
significant, resulting in a spreading of the packet. The next
stage of a long-term evolution of semiclassical wave packets for a
large class of quantum systems is universal as was shown in
Ref.\cite{A1}. Thus, at the revival time $T_R=4\pi\hbar/E''_{n_0}$
the additional phases of the individual components of the wave
packet due to quadratic terms in the expansion are exactly
multiplied by $2\pi$ which means a full reshaping of the wave
packet. At the times of fractional revivals $t=mT_R/n$, where $m$
and $n$ are mutually prime integers, the phase differences between
subsets of the component eigenfunctions are stationary, and the
wave packet breaks up into sets of sub-packets. After a few
revivals one obtains a complicated distribution of probability
density around the classical trajectory.

During some past decades the phenomena of collapse and revival
have been investigated theoretically and experimentally in
different quantum  systems.  In particular, revivals and
fractional revivals have been investigated theoretically for
Ridberg atoms, molecules and in low dimensional quantum
structures. Firstly, the detailed calculation of a short laser
pulse exitation of the high-energies Rydberg states and the total
power radiated by the atom was performed in Ref.\cite{Par}. The
space-time dynamics of the hydrogen atom was considered in
Ref.\cite{Dac} by constructing a minimum-uncertainty wave packet
that travels along a kepler orbit. It was shown that both
classical and quantum properties associated with the discreteness
and non-equidistance of the atomic spectrum  occur inevitably in
the long-term evolution of wave packets. Romera and
Santos\cite{Rom} studied the cyclotron dynamics of electrons in a
monolayer graphene, where low energy excitations are massless
Dirac fermions. They showed that when electrons are described by
wave packets with a Gaussian population of only positive energy
Landau levels, the presence of the magnetic field induces revivals
of the electron currents, besides the classical cyclotron motion.
When the population comprises both positive and negative energy
Landau levels, revivals of the electric charge current manifest
simultaneously with {\it Zitterbewegung} (ZB) and the classical
cyclotron motion. In Ref.\cite{Tor} the quasiclassical evolution
as well as the phenomenon of collapse and revival of a wave
packets in graphene quantum dots in a perpendicular magnetic field
was studied.

The Dirac equation also predicts an unexpected features of
relativistic motion. Bermudes et al.\cite{Berm} considered the
evolution of relativistic wave packets in a magnetic field and
predicted three regimes: macroscopic, microscopic and mesoscopic
(intermediate) when the electron energy level comprises several
tens of Landau levels. The authors of Ref.\cite{Berm} applied an
exact mapping between relativistic model and a combination of
Jaynes-Cummings and anti-Jaynes-Cummings interactions\cite{JC}
widely used by the quantum optics community. In the work
\cite{Berm} the relativistic version of "Schr\"odinger cat"
states, or "Dirac cat" states, was built for relativistic Landau
levels when an external magnetic field couples the states of a
relativistic spin $1/2$ charged particle.

Rusin and Zawadzki\cite{RZ} considered the effect of trembling motion of wave packet center
or ZB for relativistic electrons moving in a vacuum in the presence of an external magnetic field. As
in the original work by Schr\"odinger\cite{Sh} they used one-electron approximation and found that in this
case the effect of ZB is very small. The authors of Ref.\cite{RZ} noted, that in accordance with results
obtained by Krekora\cite{Kre}, the fully occupied electron negative energy states, or Dirac vacuum, may prevent
the interference and ZB to occur. For this reason the {\it Zitterbewegung} is merely a mathematical property
of the one-electron Dirac equation and can not be observed for real electrons. So, the calculations in
Ref.\cite{RZ} were made for 2+1 and 3+1 Dirac equations for parameters, which correspond to the motion
of trapped ions in recent experiments by Gerritsma and coworkers\cite{Ger}.

Gea-Banacloche evaluated the asymptotic description for the
evolution of two-level atom interacting with a quantized field in
initially coherent state\cite{Gea}. The collapse and revival was
investigated in this system. It was stated that collapse appears
to be associated with a measurement of the initial state of the
atom with the field. At the half-revival time the pure state of
the field, which is a mesoscopic superposition of coherent states,
is referred to as a "Schr\"odinger cat".

The phenomena of collapse and revival was observed experimentally
in different nonlinear quantum models. Among them, the two-level
atom trapped in a high-Q microwave cavity\cite{Auf}, Rydberg wave
packets in atom\cite{Yea} and molecular vibrational
states.\cite{Vra} Interestingly, methods for laser isotope
separation which take advantage of the revival phenomenon, were
proposed and  realized in Ref.\cite{A2}. In experiment Monroe et
all\cite{Mon} a superposition of two different coherent states was
created for an ion oscillating in a harmonic potential.

The quantum dynamics of free relativistic particles represented by three-dimensional
Gaussian wave packets with different initial spin polarizations was investigated and visualized in
Ref.\cite{DMPF}.
The influence of the symmetry of the initial wave packet on its kinematic characteristics such as average
electron velocity, the direction of trembling motion as well as the distribution of spin densities was
analyzed also. The effects of splitting and ZB of electron wave packets in the semiconductor quantum
well under the influence of the Rashba spin orbit coupling and also in monolayer graphene were
considered in Refs.\cite{{DMF,MDF}}.

In this work we study the long-term cyclotron dynamics of relativistic electrons, which obey the Dirac equation.
For this purpose we use the inertial frame in which the component of momentum parallel
to the direction of the magnetic field equals to zero. Such choice allows to represent
the most of results in the clear analytical form. To describe the behavior of the
relativistic particle we consider the superposition of the eigenstates of the Dirac
Hamiltonian which is centered around some quantum number $n_0\gg 1$. The bispinor structure
of electron wave function and existence of two energy bands lead to a quite complex dynamics of wave packets.
The space-time evolution of electron probability densities, as well as spin densities was investigated
analytically and numerically and after that visualized.

The paper is organized as follows. In Sec. II the relativistic
electron eigenstates and eigenvalues of Dirac Hamiltonian in a
magnetic field are introduced and the analogy with Jaynes-Cummings
model is discussed. In Sec.III we consider the dynamics of
electron wave packets when the population of Landau levels
comprises positive energy states only. The structures, which form
at fractional revival times are visualized and physical effects of
appearance of these structures are discussed.  The detailed
description of spin precession is presented. In  Sec.IV we
consider the evolution of mesoscopic wave packet, which is a
superposition of the states with positive and negative energies.
We discuss some peculiarities accompanying the revival phenomenon
for such wave packet, the effect of ZB, in particular. The results
obtained in this Section can be useful for observation of collapse
and revival in experiments with trapped ions\cite{Kre}. Finally,
in Sec. V, we conclude with the discussion of results. Some
auxiliary mathematical results are presented in Appendices A, B,
and C.

\section{Model and approach}

We consider the cyclotron dynamics of relativistic electron in a magnetic field ${\bf
B}=(0,0,B)$. Using the Landau gauge, where ${\bf A}=(-B\cdot y, 0,0)$, the Hamiltonian
can be expressed as follows:
$$
\displaylines{\hat H=c\alpha_x(\hat p_x-eBy/c)+c\alpha_y\hat p_y+\cr
\hfill +c\alpha_z\hat p_z+mc^2\beta,\hfill\llap{(1)}\cr}
$$
where $c$ is the light velocity, $m$ is the electron mass, $(-e)$ is the electron
charge and $\alpha_i$ and $\beta$ are Dirac matrices
$$
\alpha_i=\pmatrix{0&\sigma_i \cr
\sigma_i &0},\quad \beta=\pmatrix{I&0\cr 0& -I\cr}, \qquad i=1,2,3,
\eqno{(2)}
$$
$\sigma_i$- are Pauli matrices.
The Dirac equation admits a solution of the type
$$
\psi({\bf r})=\frac{\exp\Big(\frac{ip_xx+ip_zz}{\hbar}\Big)}{2\pi\hbar}U(y).\eqno{(3)}
$$
For simplicity, we assume in Eqs.(1) and (3), $p_z=0$ (we always may choose the
inertial frame, in which this condition takes place). Then the equation that we consider is
$$
\hat H U=EU,\eqno{(4)}
$$
where the effective Hamiltonian $\hat H$
$$
\hat H=\pmatrix{mc^2 & c\hat p_y\sigma_y-eB(y-y_c)\sigma_x\cr
c\hat p_y\sigma_y-eB(y-y_c)\sigma_x & -mc^2\cr},\eqno{(5)}
$$
and $y_c=cp_x/eB$.
It is easy to check that the operator
$$
\hat K=\pmatrix{\sigma_z & 0\cr
0 & -\sigma_z\cr},\eqno{(6)}
$$
commutes with Hamiltonian (5) and consequently we can construct a function which
is an eigenfunction of $\hat H$ and $\hat K$ simultaneously. As follows from
Eq.(6), the eigenstates of $\hat K$ with eigenvalues $\lambda_k=\pm 1$ can be written as
$$
\psi_{\lambda_k=1}=(f|\uparrow >,\chi |\downarrow >)^T=(f,0,0,\chi)^T,\eqno{(7a)}
$$
$$
\psi_{\lambda_k=-1}=(f|\downarrow >,\chi |\uparrow >)^T=(0,f,\chi ,0)^T,\eqno{(7b)}
$$
where $f=f({\bf r},t)$, $\chi=\chi({\bf r},t)$ are arbitrary functions of ${\bf r}$
and $t$. Spin-up $|\uparrow >$ and spin-down $|\downarrow >$ spinors are given by
$$
|\uparrow >=\pmatrix{1\cr 0\cr},\quad |\downarrow >=\pmatrix{0\cr 1\cr}.\eqno{(8)}
$$
We see that for a given value $\lambda_k$ two components of the Dirac spinors (7a),
(7b) are vanished: $\psi_2=\psi_3=0$ for $\lambda_k=1$ and $\psi_1=\psi_4=0$ for
$\lambda_k=-1$. Thus, depending on the sign of $\lambda_k$, the Hilbert space of
solutions of the Dirac equation (4) with Hamiltonian Eq.(5) can be divided into two
invariant subspaces. Correspondingly, the Dirac Hamiltonian is decomposed into two
terms. One of them couples the components $\psi_1$ and $\psi_4$, and this part is identical
to a Jaynes-Cummings model, describing the interaction of two-level atom with a
quantized single-mode field.\cite{JC} The remaining term which couples the components
$\psi_2$ and $\psi_3$ is identical to the anti-Jaynes-Cummings interaction.\cite{Berm}
This suggests that we can deal with two-component wave function
$\psi(y)=(f(y),\chi(y))^T$. For $\lambda_k=1$ one obtains from Eqs.(4), (5) and (7a)
$$
\pmatrix{mc^2 & -\sqrt{2\hbar ecB}\hat a\cr
-\sqrt{2\hbar ecB}\hat a^{\dagger} & -mc^2\cr}\pmatrix{f(y)\cr \chi(y) \cr}=
E\pmatrix{f(y)\cr \chi(y) \cr}.\eqno{(9)}
$$
Here we introduce the raising and lowering operators for the harmonic oscillator
$$
\hat a^{\dagger}=\frac{1}{\sqrt{2}}\bigg(\sqrt{\frac{m\omega}{\hbar}}(y-y_c)-
\frac{i\hat p_y}{\sqrt{\hbar m\omega}}\bigg),\eqno{(10a)}
$$
$$
\hat a=\frac{1}{\sqrt{2}}\bigg(\sqrt{\frac{m\omega}{\hbar}}(y-y_c)+
\frac{i\hat p_y}{\sqrt{\hbar m\omega}}\bigg),\eqno{(10b)}
$$
where $\omega=eB/mc$ is the cyclotron frequency for a non-relativistic electron.
The eigenvalues of Eq.(9) are:
$$
E_{s,n}=s\cdot\varepsilon_n=s\sqrt{m^2c^4+2\hbar eBcn}\eqno{(11)}
$$
with $n=1,2,3,\ldots$ for $s=1$ and $n=0,1,2,3,\ldots$ for $s=-1$. Correspondent
eigenstates are given by
$$
\psi_{\lambda_k=1,s=1,n}(y)=\pmatrix{d_n\phi_{n-1}(y-y_c)\cr
-b_n\phi_n(y-y_c)\cr},\eqno{(12)}
$$
$$
\psi_{\lambda_k=1,s=-1,n}(y)=\pmatrix{b_n\phi_{n-1}(y-y_c)\cr
d_n\phi_n(y-y_c)\cr},\eqno{(13)}
$$
where $\phi_n(y-y_c)$ is linear oscillator wave function, and the coefficients $d_n$ and
$b_n$ are determined by
$$
d_n=\sqrt{\frac{\varepsilon_n+mc^2}{2\varepsilon_n}},
\quad b_n=\sqrt{\frac{\varepsilon_n-mc^2}{2\varepsilon_n}}.\eqno{(14)}
$$
Two-component wave function for $\lambda_k=-1$ obeys the Dirac equation which is
similar to Eq.(9):
$$
\pmatrix{mc^2 & -\sqrt{2\hbar ecB}\hat a^{\dagger}\cr
-\sqrt{2\hbar ecB}\hat a & -mc^2\cr}\pmatrix{f(y)\cr \chi(y) \cr}=
E\pmatrix{f(y)\cr \chi(y) \cr}.\eqno{(15)}
$$
As before, the energies $E_{s,n}$ are determined by Eq.(11), but now
$n=0,1,2,3,\ldots$ for $s=1$ and $n=1,2,3,\ldots$ for $s=-1$. The associated
eigenstates are
$$
\psi_{\lambda_k=-1,s=1,n}(y)=\pmatrix{d_n\phi_n(y-y_c)\cr
-b_n\phi_{n-1}(y-y_c)\cr},\eqno{(16)}
$$
$$
\psi_{\lambda_k=-1,s=-1,n}(y)=\pmatrix{b_n\phi_n(y-y_c)\cr
d_n\phi_{n-1}(y-y_c)\cr}.\eqno{(17)}
$$
Note that in the considered case $p_z=0$ the four-component wave function $U(y)$ in
Eq.(3) is connected with two-component wave function $\psi_{\lambda,s,n}(y)$ (see Eqs.(12),
(13), (16) and (17)) by relations
$$
U_{\lambda_k=1,s,n}(y)=\pmatrix{|\uparrow > & 0 \cr
0 & |\downarrow > \cr}\psi_{\lambda=1,s,n}(y),\eqno{(18)}
$$
$$
U_{\lambda_k=-1,s,n}(y)=\pmatrix{|\downarrow > & 0 \cr
0 & |\uparrow > \cr}\psi_{\lambda=-1,s,n}(y),\eqno{(19)}
$$
Thus, in the considered case $(p_z=0)$ the general solution can be written as
$$
\psi({\bf r},t)=\int dp\, \varphi_p(x)\sum_{n,s,\lambda_k}C_{n,\lambda_k}^{s}(p)
U_{\lambda_k,s,n}(y)\exp(-is\varepsilon_nt/\hbar),
\eqno{(20)}
$$
where $\varphi_p(x)=1/\sqrt{2\pi\hbar}\exp(ipx/\hbar)$, and coefficients
$C_{n,\lambda_k}^{s}(p)$ are to be determined by the initial wave function.

\section{Evolution of wave packet formed by the positive energy eigenstates}

\subsection{Time dependence of the electron probability density}

To study the complex dynamics of a real relativistic electron
placed in the magnetic field, we construct the initial mesoscopic
wave packet as a linear combination of the positive energy states
only. Besides, we consider the special form of the initial wave
function which represents the coherent state of the
non-relativistic electron in the magnetic field. So, let the
coefficients $C_{n,\lambda_k}^{s}(p)$ in Eq.(20) are determined by
the expressions:
$$
\displaylines{
C_{n\geq 0,\lambda_k=-1}^{s=1}(p)=\frac{\beta g(p)c_{n+1}}
{\sqrt{\alpha^2+\beta^2}},\qquad C_{n\geq 1,\lambda_k=1}^{s=1}(p)=
\frac{\alpha g(p)c_n}{\sqrt{\alpha^2+\beta^2}}\cr
\hfill C_{n,\lambda_k}^{s=-1}(p)=0,\hfill\llap{(21)}\cr}
$$
where
$$
g(p)=\sqrt{\frac{a}{\hbar\sqrt{\pi}}}\exp\Big(-\frac{1}{2}(pa/\hbar-qa)^2\Big),
\eqno{(22)}
$$
$$
c_n=\frac{\exp(-(qa)^2/4)\cdot (-qa)^{n-1}}{\sqrt{2^{n-1}(n-1)!}},\eqno{(23)}
$$
$a=\sqrt{\hbar c/eB}$ is the magnetic length, the parameter
$qa\sim\sqrt{n_0}$ characterizes the radius of relativistic orbit.
For simplicity we assume $\alpha=\alpha^{\star}$,
$\beta=\beta^{\star}$. Then from Eq.(20) we have
$$
\displaylines{\psi({\bf r},t)=\frac{1}{\sqrt{\alpha^2+\beta^2}}\int dp \, \varphi_p(x)
g(p)\times\cr
\hfill\Bigg[\sum_{n=1}\pmatrix{d_n(\alpha c_n\phi_{n-1}(y)|\uparrow >+
\beta c_{n+1}\phi_{n}(y)|\downarrow >)\cr
-b_n(\alpha c_n\phi_{n}(y)|\downarrow >+\beta c_{n+1}\phi_{n-1}(y)|\uparrow >)\cr}
e^{-i\varepsilon_nt/\hbar}+ \hfill\cr
\hfill \beta c_1 \pmatrix{\phi_0(y)|\downarrow>\cr 0\cr}e^{-i\varepsilon_0t/\hbar}
\Bigg].\hfill\llap{(24)}\cr}
$$
Note that in the considered case all components of the wave
function differ from zero. Thus, for describing the wave packet
dynamics we should use the full Hamiltonian, Eq.(1). To analyze
the motion of the packet center we first of all have to find the
average value of the velocity operator $\hat v_i=c\alpha_i$. Using
Eq.(2) and Eqs.(22), (23), (24) we arrive after integration over
$x$ and $y$ at the following equation:
$$
\displaylines
{
\overline{v_x}(t)+i\overline{v_y}(t)=\frac{ce^{-(qa)^2/2}}{\alpha^2+\beta^2}
\sum_{n=0}\frac{(qa)^{2n+1}}{2^n
n!}\sqrt{\frac{\varphi_{n+1}-1}{2(n+1)\varphi_{n+1}}}\times\cr
\hfill\Bigg(\alpha^2\sqrt{\frac{\varphi_{n+2}+1}{\varphi_{n+2}}}e^{i(\varphi_{n+2}-
\varphi_{n+1})\tau}+\beta^2\sqrt{\frac{\varphi_n+1}{\varphi_n}}e^{i(\varphi_{n+1}-
\varphi_n)\tau}\Bigg).\hfill\llap{(25)}\cr
}
$$
Here and below $\varphi_n$ is the electron energy (in units $mc^2$)
$$
\varphi_n=\sqrt{1+2n(\lambda/a)^2},\eqno{(26)}
$$
and $\tau=ct/\lambda$, where $\lambda=\hbar/mc$ is the Compton wave length. It can be
shown that the dominant contribution to the sum in Eq.(25) comes from the interval in
the neighborhood of $n\approx n_0=(qa)^2/2$. As follows from Eq.(26) for
$n(\lambda /a)^2\ll 1$, $\varphi_n\approx 1+n(\lambda /a)^2$ and Eq.(25) describes the
cyclotron motion of a non-relativistic electron
$$
\displaylines
{
\overline{v_x}(t)=\frac{\hbar q}{m}\cos\omega t,\cr
\hfill\overline{v_y}(t)=\frac{\hbar q}{m}\sin\omega t.\hfill\llap{(27)}\cr
}
$$

\begin{figure}
  \centering
  \includegraphics[width=150mm]{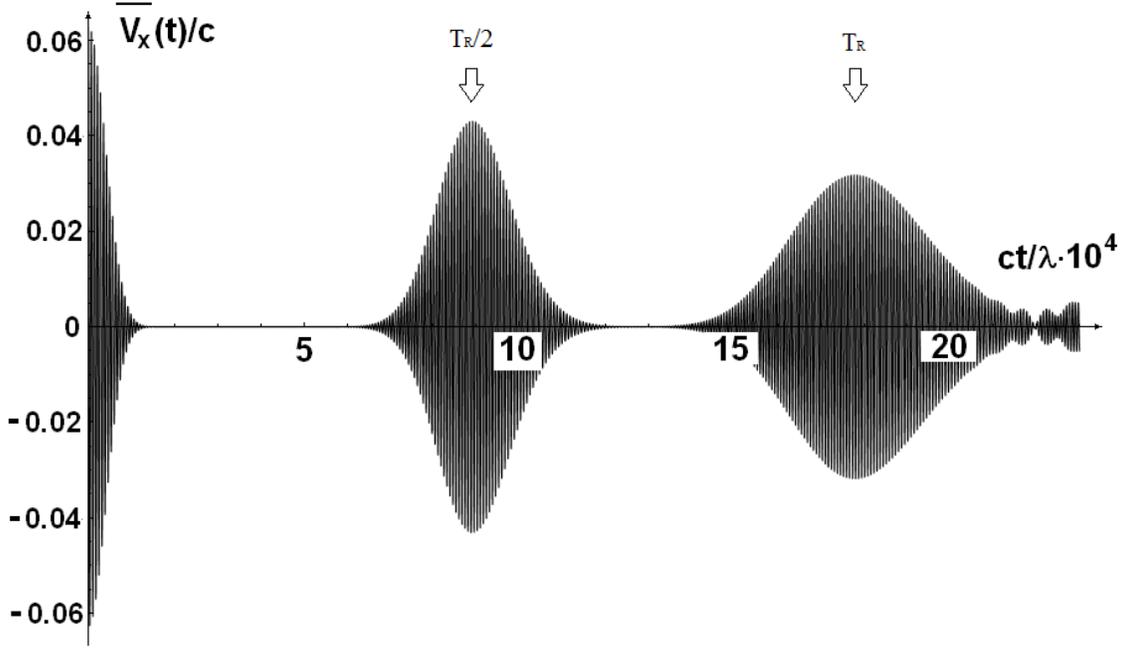}
  \caption{The dependence of the average velocity $\overline{v_x}(t)$ versus time
for the initial wave packet, Eq.(24), with $\alpha=\beta$ and the
parameters $\lambda/a=0.1$, $qa=5$ (that correspond to the value
$n=n_0\approx 13$ and dispersion $\triangle
n\sim\sqrt{n_0}\approx3.5$).}
\end{figure}

If $n_0$ is large enough, the quantum interference of the wave
packet components leads to the different types of periodicity. In
Fig.1 we plot the dependence of $\overline{v_x}(t)$  for a
relativistic wave packet with $qa=5$, $\alpha =\beta$, placed in
the magnetic field $B\simeq 4.5\cdot 10^7$T. (Note that this
magnetic field is 1.5 times greater than the maximum value
achieved currently in laboratory experiments). In this case
$\varphi_{n+1}-\varphi_n\approx\varphi '_{n_0}+(\varphi
''_{n_0}/2)(2n-2n_0+1)$. So, in the beginning the average velocity
of the wave packet center oscillates with classical cyclotron
motion period
$$
T_{cl}=\frac{2\pi\hbar}{mc^2\varphi '_{n_0}}=\frac{2\pi\varphi_{n_0}}{\omega_c},\;
\omega_c=\frac{eBc}{\varepsilon_{n_0}}.
\eqno{(28)}
$$
At large times $t>T_D$ these oscillations fade away but then appear again. The damping
time $T_D$ can be estimated from Eq.(A4). For $|\varphi ''_{n_0}|\tau\ll 1$ the
amplitude of the oscillations is proportional to $\exp(-(qa\varphi ''_{n_0}\tau /2)^2)=
\exp(-t^2/T_D^2)$, where
$$
T_D=\frac{2\lambda}{qac|\varphi ''_{n_0}|}=\frac{2\hbar}{qa|\varepsilon ''_{n_0}|}.
\eqno{(29)}
$$
At times $T_D<t<T_R/2$ the average velocity components as well as
average coordinates of the packet center are negligible. At $t\sim
T_R/2$ (see Fig.1) the velocity oscillations will occur again.
Here $T_R$ is the revival time at which the wave packet is fully
reformed.\cite{A1} At $t_k=kT_R$ where $k$ is integer, the
nonlinear (quadratic) term in the expansion of $\varphi_n$ has no
effect, so that
$$
T_R=\frac{4\pi\lambda}{c|\varphi ''_{n_0}|}=\frac{4\pi\hbar}{|\varepsilon ''_{n_0}|}.
\eqno{(30)}
$$
Note that according to Averbukh and Perel'man\cite{A1} the moment $T_R/2$ is the first
reconstruction time of the wave packet (see more detailed discussion below). For the
considered wave packet Eqs.(29) and (30) give $T_R/T_D\sim 30$ that is in a good
agreement with numerical calculations (Fig.1).

Now we describe some peculiarities of the space-time dynamics of the electron wave packet
(24). Performing the integration over $p$ in Eq.(24), one has
$$
\displaylines{
\psi({\bf r},t)=\frac{M(\rho,\theta)}{\sqrt{\alpha^2+\beta^2}}\sum_{n=0}\bigg[
\frac{\gamma^{n-1}}{(n-1)!}(1-\delta_{n,0})
e^{i\theta}\Big(\alpha d_n, 0, \frac{\beta b_n qa}{\sqrt{2n}}, -\frac{\alpha b_n\rho}
{\sqrt{2n}}e^{-i\theta}\Big)^T +\cr
\hfill \frac{\gamma^n}{n!}\beta d_n (0,1,0,0)^T\bigg]e^{-i\theta
n-i\varphi_nct/\lambda}.\hfill\llap{(31)}\cr}
$$

\begin{figure}
\begin{center}
(a)\includegraphics[width=0.4\textwidth]{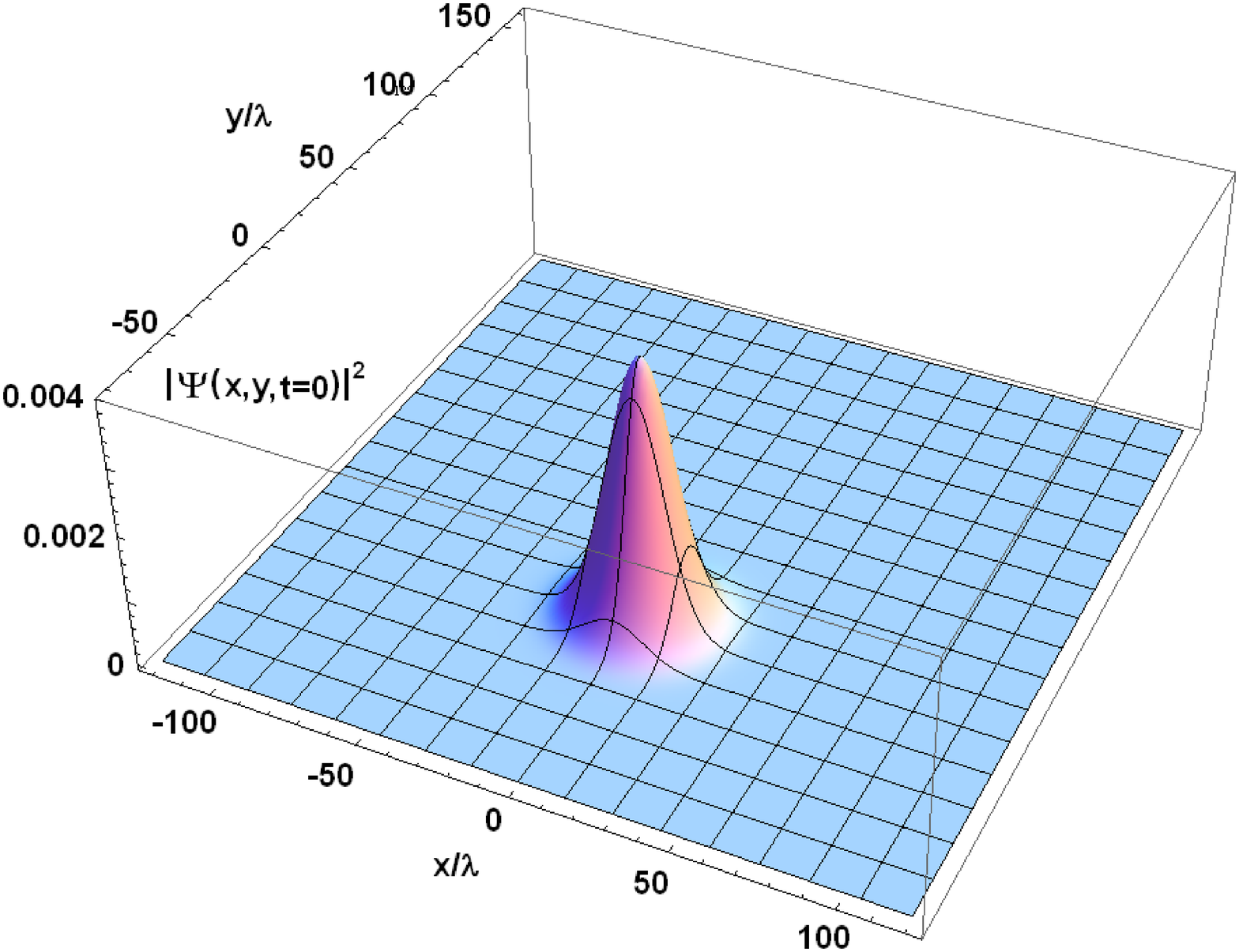}
(b)\includegraphics[width=0.4\textwidth]{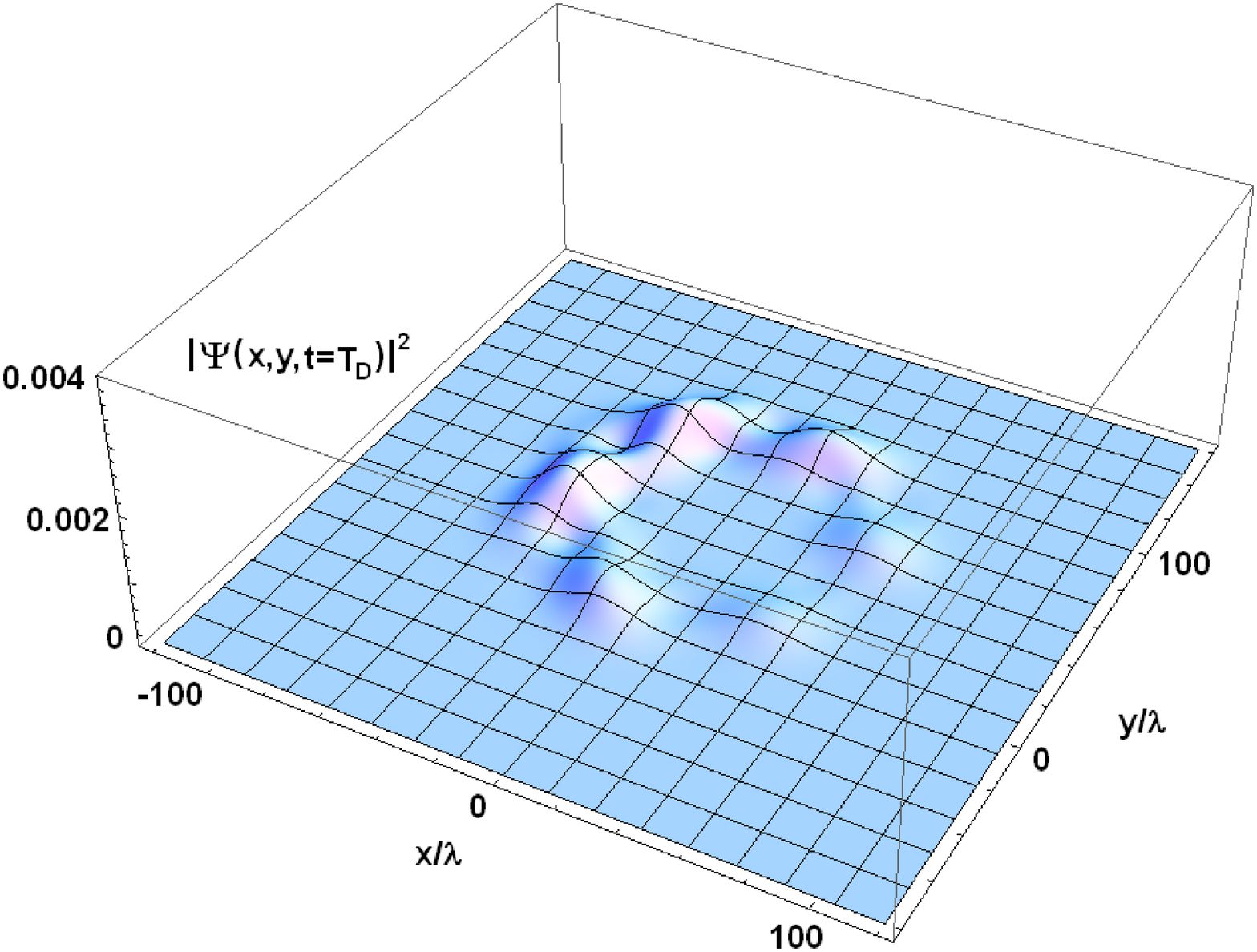}
\\
(c)\includegraphics[width=0.4\textwidth]{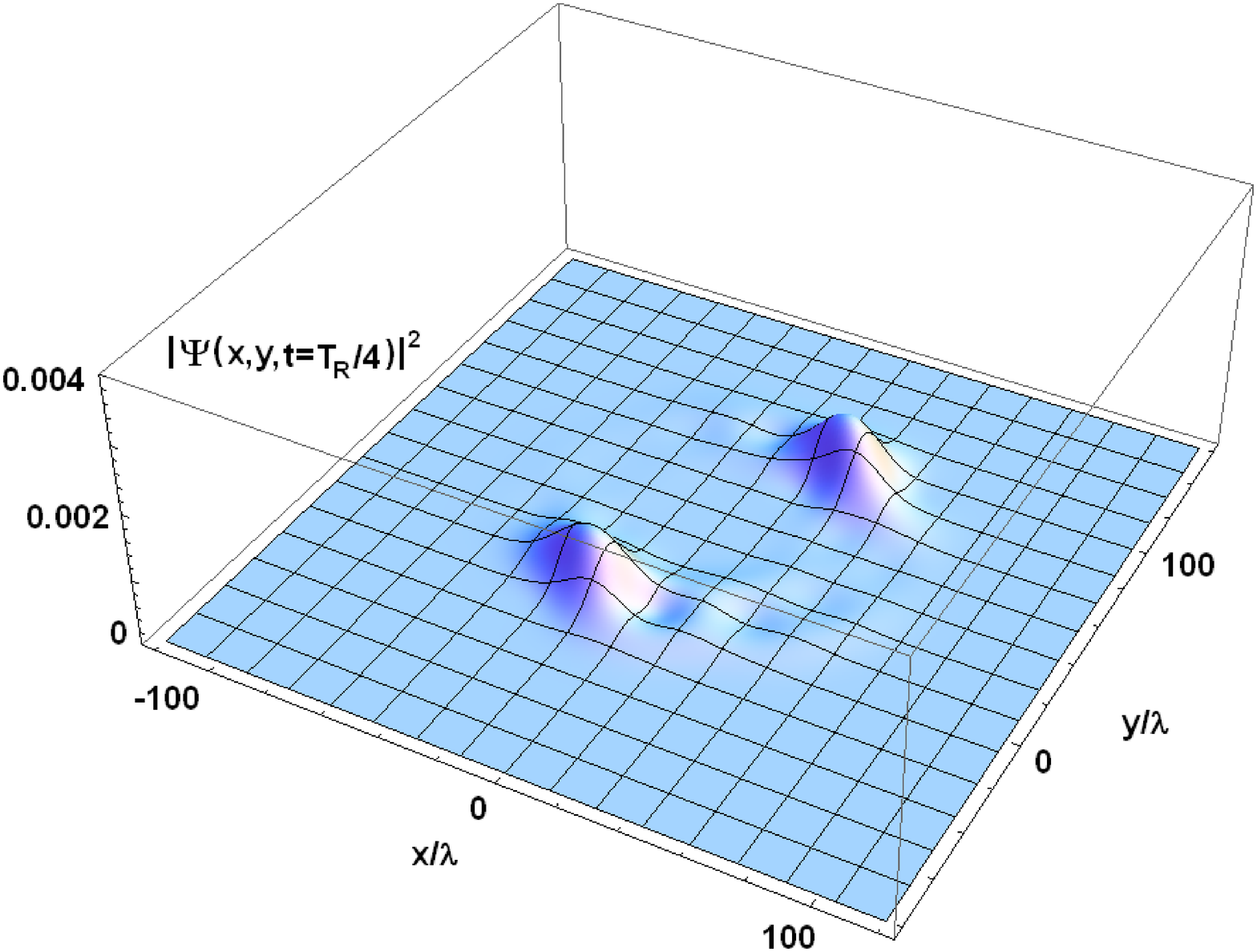}
(d)\includegraphics[width=0.4\textwidth]{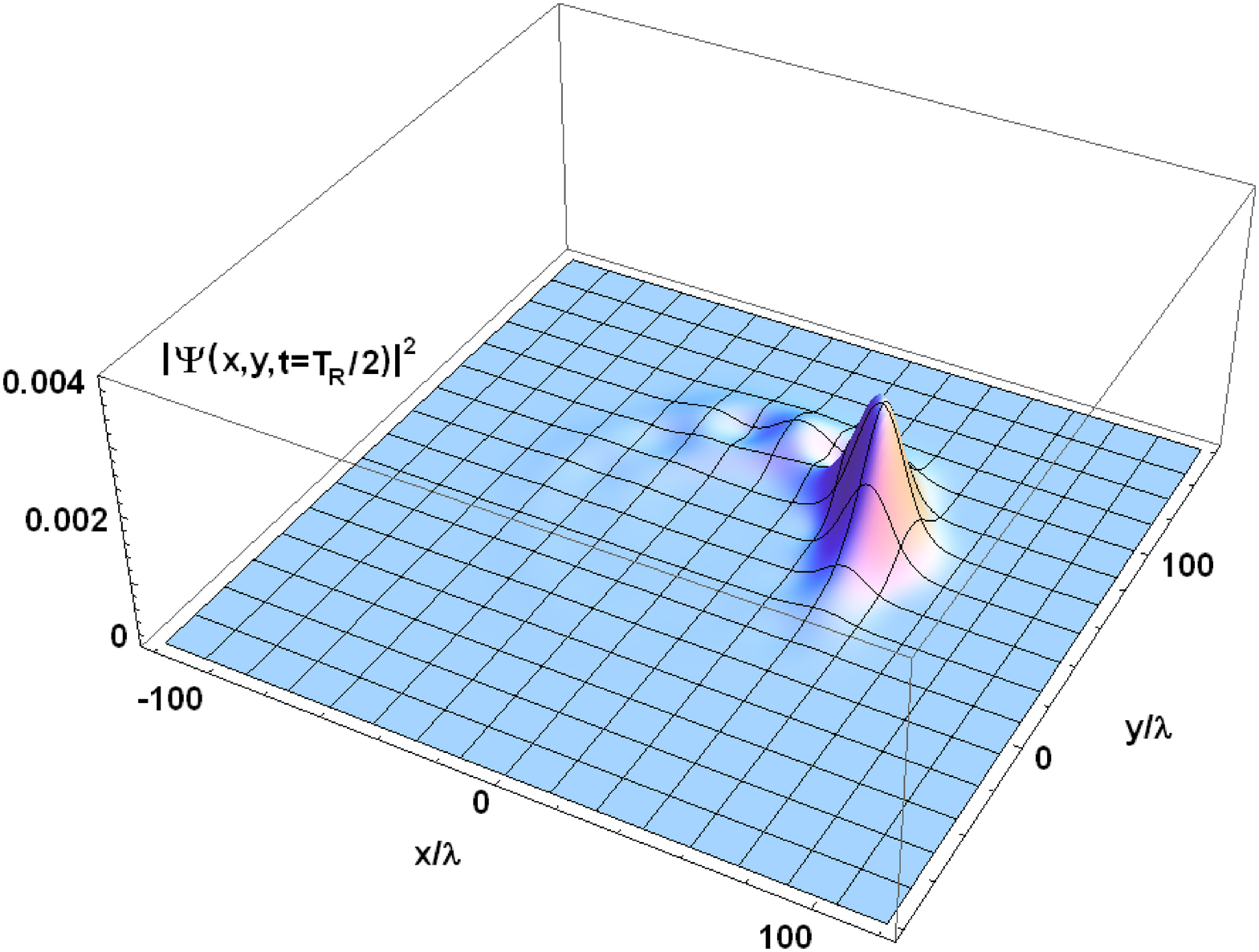}
(e)\includegraphics[width=0.4\textwidth]{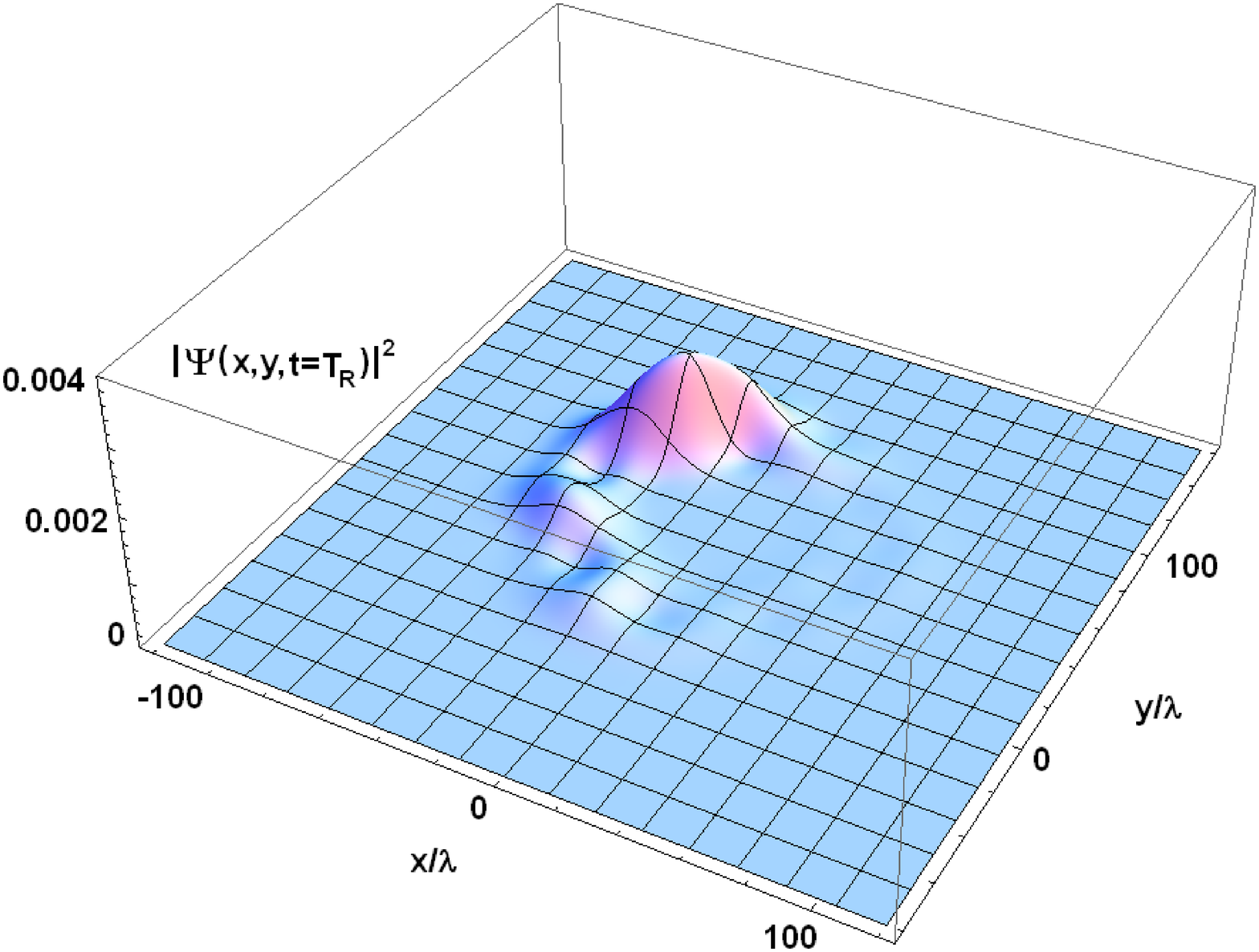} \caption{
(Color online) The probability densities for the wave packet,
Eq.(24), with $ \alpha=\beta$ and the parameters $\lambda/a=0.1$,
$qa=5$ at times: (a) $t=0$, (b) $t\approx T_D$, (c) $t\approx
T_R/4$, (d) $t\approx T_R/2$, (e) $t\approx T_R$. }
\end{center}
\end{figure}

The details of the calculation are presented in the Appendix B. Here we use the
notations
$$
\frac{x}{a}=\rho\sin\theta,\, \frac{y-qa^2}{a}=\rho\cos\theta,\,
\gamma=-\frac{qa\rho}{2}, \eqno{(32)}
$$
$$
M(\rho,\theta)=\frac{\exp\Big(-\frac{\rho^2+(qa)^2}{4}+\frac{i\rho\sin\theta
(\rho\cos\theta+2qa)}{2}\Big)}{a\sqrt{2\pi }}.
\eqno{(33)}
$$

In Fig.2 we plot the time evolution of the probability density $|\psi({\bf r},t)|^2$
calculated with the use of Eqs.(31), (32) and (33).

At the initial stage $t\ll T_D$ the motion of the wave packet is determined by the linear
term of the expansion (A2). In this approximation
$$
\psi({\bf r},t)=\psi_{cl}({\bf r},t),
\eqno{(34)}
$$
where $\psi_{cl}({\bf r},t)$ is determined by Eq.(B4). As was
noted above in our case the value $2n_0(\lambda/a)^2\simeq
(q\lambda)^2=0.25$ so that $\psi_{cl}({\bf r},t)$ differs slightly
from the coherent wave function of non-relativistic electron (see
Eq.(B4)). Therefore, the shape of the wave packet $|\psi_{cl}({\bf
r},t)|^2$ remains unchanged during its cyclotron motion, Eq.(B6),
with the classical period $T_{cl}$ of the order of attoseconds
($10^{-18} s.$).

However, after a few orbits at $t\sim T_D$ (Fig.2b) the quadratic term in the expansion (A2)
becomes significant, that leads to the dephasing of different terms in the
superposition Eq.(31). As a result, the wave packet is decomposed in a host of
sub-packets. As was shown by Averbukh and Perel'man\cite{A1} at time $t\approx mT_R/n$,
where $m$ and $n$ are integer and $m/n$ is an irreducible fraction, the number of such
sub-packets is equal to $N=n(3-(-1)^n)/4$. The first reconstruction of wave packet
takes place around time $t_1=T_R/2$\cite{A1} where the revival time $T_R$ is
determined by Eq.(30) and $T_R\approx 2.3\cdot 10^{-16}\, s$ in our case. At that time
$t_1$ the $n-$th term in the superposition (31) acquires the phase factor $f_k=e^{i\pi
k^2}=e^{i\pi k}$, $k=n-n_0$, that leads to the result for the wave packet function
(see Fig.2d)
$$
\psi({\bf r},t_1)=\psi_{cl}\Big({\bf r},t_1+\frac{T_{cl}}{2}\Big).
\eqno{(35)}
$$
It should be noted in accordance with Eq.(B7) that the position of
the reconstructed packet at the moments $T_R/2$ and $T_R$ does not
coincide with initial position of the packet at $t=0$. At early
times, for example, at $t_2=T_R/4$, the original wave packet
splits into two subpackets (the fractional revival). In fact, the
additional phase factor in this case $f_k=e^{i\pi k^2/2}$ is a
periodic function $f_k=f_{k+2}$, so one can use the ordinary
Fourier expansion\cite{A1}
$$
f_k=p_0+p_1e^{i\pi k},\, k=0,1, \eqno{(36)}
$$
from which
$$
p_0=\frac{e^{i\pi/4}}{\sqrt{2}},\,
p_1=\frac{e^{-i\pi/4}}{\sqrt{2}}. \eqno{(37)}
$$
Then, by substituting Eqs.(36) and (37) into Eq.(33) we obtain
$$
\psi({\bf r},t_2=T_R/4)=\frac{e^{i\pi/4}}{\sqrt{2}}\psi_{cl}({\bf r},t_2)+
\frac{e^{-i\pi/4}}{\sqrt{2}}\psi_{cl}({\bf r},t_2+T_{cl}/2).
\eqno{(38)}
$$
Note that according to Eqs.(B7) and (38), the angular distance
between these two packets is $180^{\circ}$ (Fig.2c). Similar
structure consisting of $N$ sub-packets evenly distributed along
the classical orbit arises at the other fractional revival times.
It should be noted, however, that the effects of fractional
revivals are not manifested in the dependence of the mean velocity
on the time (see Fig.1). The shape of the wave packet at revival
time $T_R$ (Fig.2e) differs from the initial one (Fig.2a). The
spreading of the packet at $t=T_R$ is connected with the next
(cubic) term in the Taylor series.

Features of quantum dynamics of the relativistic wave packets
described above should be seen in the nature of electromagnetic
radiation emitted by the moving electrons. The radiation field as
well as its intensity are determined by the multipole moments of
system. Thus, at the initial stage of the evolution of wave packet
($t<T_D$), the dipole radiation with a frequency
$\omega=\omega_c=\frac{eBc}{\varepsilon_{n_0}}$ is dominated. Its
intensity (in the classical approach) is proportional to the
square of the second time derivative of the average dipole moment
$\ddot{\overline{\bf d}}(t)=-e \dot{\overline{\bf v}} (t)$, where
the components of average velocity are determined by Eq. (25). At
times $T_D<t<T_R/2$ the mean velocity of the packet is negligible
(Fig.1). During this time electromagnetic emission is determined
by both the time-dependent components of the quadrupole moment
tensor $\overline{D_{\alpha,\beta}}=\int |\Psi({\bf
r},t)|^2(3x_{\alpha}x_{\beta}-{\bf r}^2\delta_{\alpha,\beta})d{\bf
r}$, and the magnetic moment $\overline{\bf m}=\frac{1}{2c} \int
\Psi({\bf r},t)^{\dagger}[{\bf r},\hat{{\bf v}}] \Psi({\bf
r},t)d{\bf r}$, where the wave function $\Psi({\bf r},t)$ is
determined be the Eqs. (31)-(33). In particular,  at the time
moment $t=T_R/4$, when the initial wave packet splits into two
sub-packets, the quadrupole radiation at the doubled frequency $2
\omega_c$ is dominated, that is confirmed by numerical
calculations of radiation intensity along $\bf n$ direction:
$dI/d\Omega=\frac{1}{144 \pi c^5} [\overline{\dddot{{\bf D}}},{\bf
n}]^2$. In the time interval $\triangle t\sim 2T_D$ near the
moment $t\sim T_R/2$ (see Fig.1) the dipole radiation begins to
dominate again. Then at large times the multipole and
magnetic-dipole terms become significant, and so on.

\subsection{Spin dynamics}

Below we describe and visualize the spin dynamics of the electron wave packet, Eq.(24), in
the presence of a magnetic field ${\bf B}=(0,0,B)$. It is well-known that in
non-relativistic quantum mechanics the electron spin precesses in the $(x,y)$-plane
with cyclotron frequency $\omega=eB/mc$, so that the $S_z$-component remains constant.
The spin operators $\hat{\bf S}$ for the Dirac particle
$$
\hat S_i=\frac{\hbar}{2}{\Sigma_i}=\frac{\hbar}{2}\pmatrix{\sigma_i & 0\cr
0 & \sigma_i\cr},\qquad i=1,2,3.
\eqno{(39)}
$$
do not commute with the Dirac Hamiltonian, Eq.(1). It follows that besides the usual spin
precession which is more complicated in the relativistic case, the $S_z$-component
changes with time in general. However, the average value of component $S_z$ is conserved for
the wave packet  consisting of the states with $E_n>0$ or $E_n<0$ only. It directly
follows from the fact the commutator $[\hat S_z,\hat H]$ is equal to zero in the subspace of the eigenvectors of
the Dirac Hamiltonian for the same sign of energy. The original wave packet (Eq.(24))
satisfies this condition. Thus, we are interested to find the time-dependent average
values of $\overline{S_x}(t)$ and $\overline{S_y}(t)$
$$
\overline{S_{x,y}}(t)=\frac{\hbar}{2}\int\, d{\bf r}\psi^{\dagger}({\bf r},t)
\Sigma_{x,y}\psi({\bf r},t),
\eqno{(40)}
$$
where $\psi({\bf r},t)$ is given by Eq.(24). It is convenient to represent the result
of calculations in the complex form
$$
\displaylines{\overline{\Sigma_x}(t)+i\overline{\Sigma_y}(t)=\frac{\alpha\beta}{\alpha^2+\beta^2}
e^{-(qa)^2/2}\sum_{n=0}\frac{((qa)^2/2)^n}{n!}\times\cr
\hfill\bigg(\sqrt{\frac{(\varphi_n+1)(\varphi_{n+1}+1)}{\varphi_n\varphi_{n+1}}}+
\sqrt{\frac{n(\varphi_n-1)(\varphi_{n+1}-1)}{(n+1)\varphi_n\varphi_{n+1}}}\bigg)
e^{ict(\varphi_{n+1}-\varphi_n)/\lambda}.\hfill\llap{(41)}\cr
}
$$

\begin{figure}
  \centering
  \includegraphics[width=150mm]{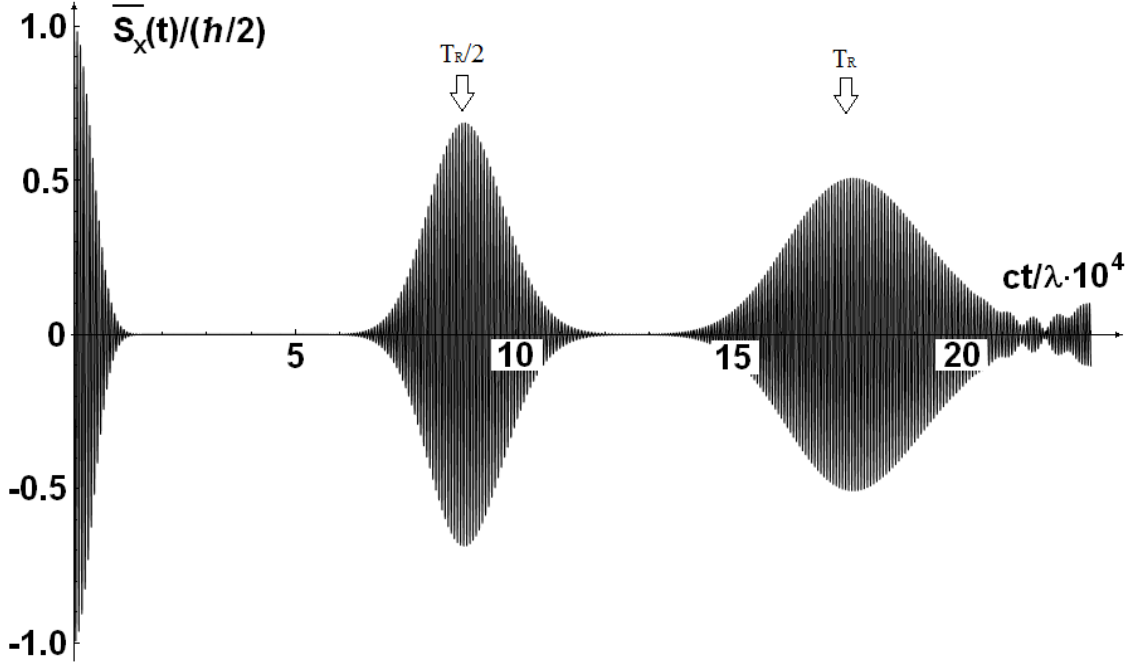}
  \caption{Time dependence $\overline{S_x}(t)$ for electron wave packet, Eq.(24), with the parameters
$\lambda/a=0.1$, $qa=5$.}
\end{figure}

The dependence $\overline{S_x}(t)$ represented in Fig.3 is similar to the behavior of the average
velocity component $\overline{v_x}(t)$ (see Fig.1). Thus,
we arrive at the conclusion that the usual precession continues during several periods
$T_{cl}$ only. At time $t\lesssim T_D$ the spin-rotation stops and arises again later at
$t\lesssim T_R/2$.
Note that the spin precession of a moving electron in a uniform magnetic field is closely
connected to the conservation of helicity  $\Sigma_i\cdot({\bf p}+e{\bf A}/c)_i$. Notice
also that, as it follows from Eq.(41), the average values of spin components
$\overline{\Sigma_x}=\overline{\Sigma_y}=0$ if the electron wave packet is a
superposition of the states with $\lambda_k=1$ (or $\lambda_k=-1$) only (see Sec.I).

Now we consider the spin densities which we denote as $<S_{x,y}({\bf r},t)>=
\frac{\hbar}{2}<\Sigma_{x,y}({\bf r},t)>$, where
$$
<\Sigma_{x,y}({\bf r},t)>=\psi^{\dagger}({\bf r},t)\Sigma_{x,y}\psi({\bf r},t).
\eqno{(42)}
$$
Using the expression for the wave function, Eq.(24), we find from Eq.(42)
$<\Sigma_x({\bf r},t)>$ and $<\Sigma_y({\bf r},t)>$ as the real and imaginary parts of
$$
\displaylines{<\Sigma_x(\rho,\theta , t)>+i<\Sigma_y(\rho,\theta , t)>=
\frac{\alpha\beta e^{-((qa)^2+\rho^2)/2}}{2\pi a^2(\alpha^2+\beta^2)}
\sum_{m,n=0}\frac{(-qa\rho/2)^{n+m}}{n! m!}\times\cr
\hfill\bigg(\sqrt{\frac{(\varphi_{m+1}+1)(\varphi_n+1)}{\varphi_n\varphi_{m+1}}}+
\sqrt{\frac{n(\varphi_n-1)(\varphi_{m+1}-1)}{(m+1)\varphi_n\varphi_{m+1}}}\bigg)\times\hfill\cr
\hfill \exp(ict(\varphi_{m+1}-\varphi_n)/\lambda+i\theta (m-n)).\hfill\llap{(43)}\cr
}
$$

\begin{figure}
\begin{center}
(a)\includegraphics[width=0.4\textwidth]{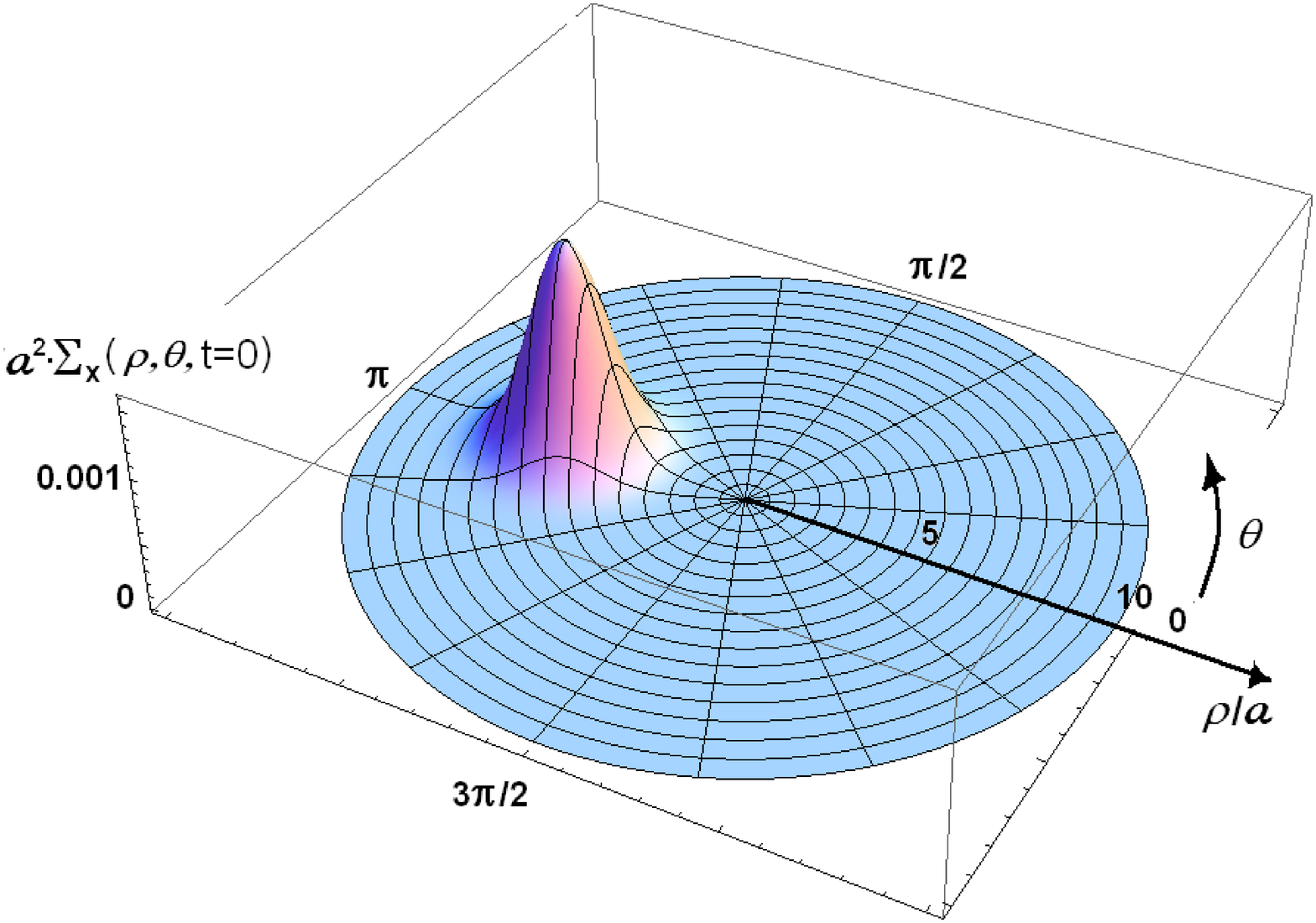}
(b)\includegraphics[width=0.4\textwidth]{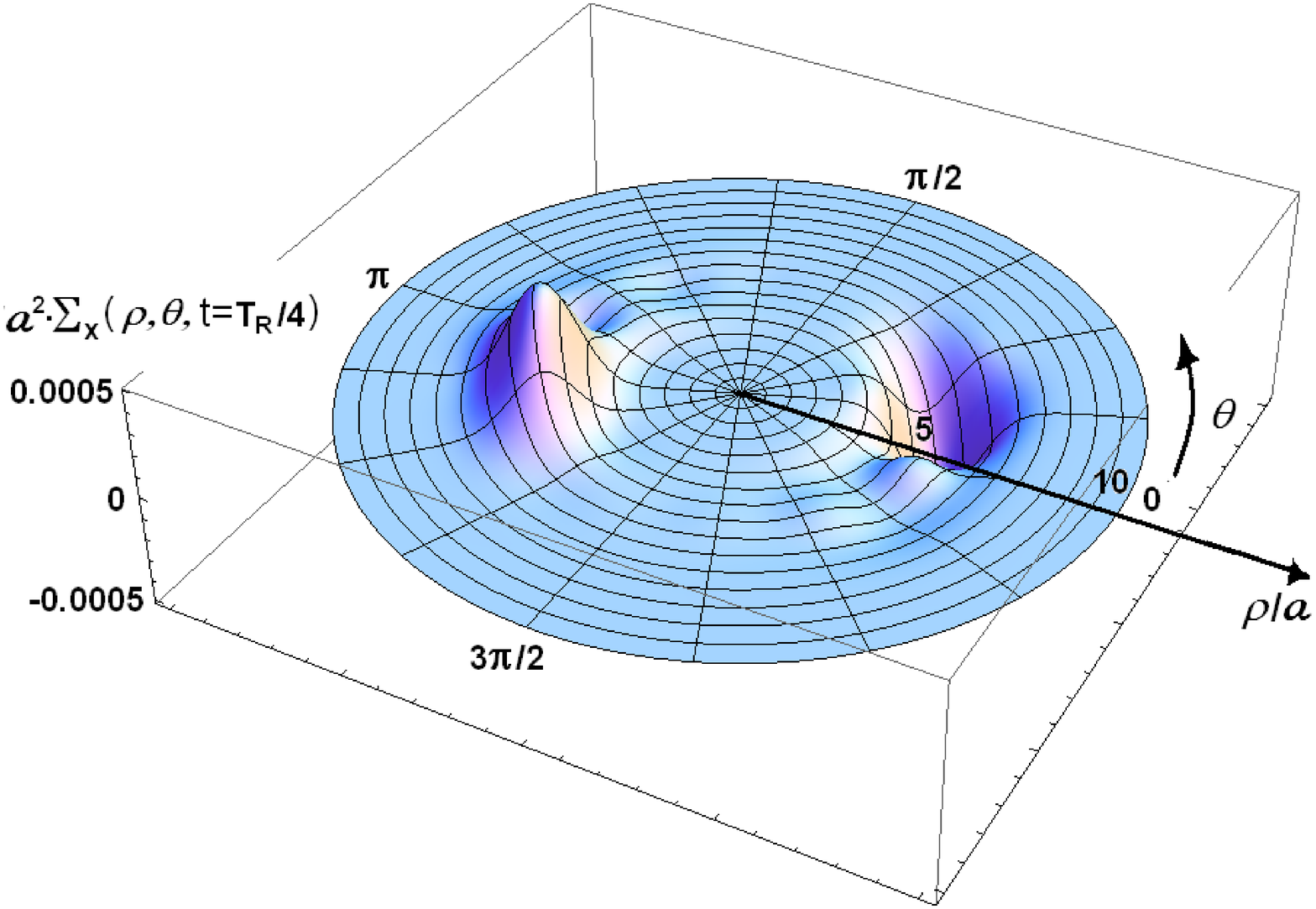}
\\
(c)\includegraphics[width=0.4\textwidth]{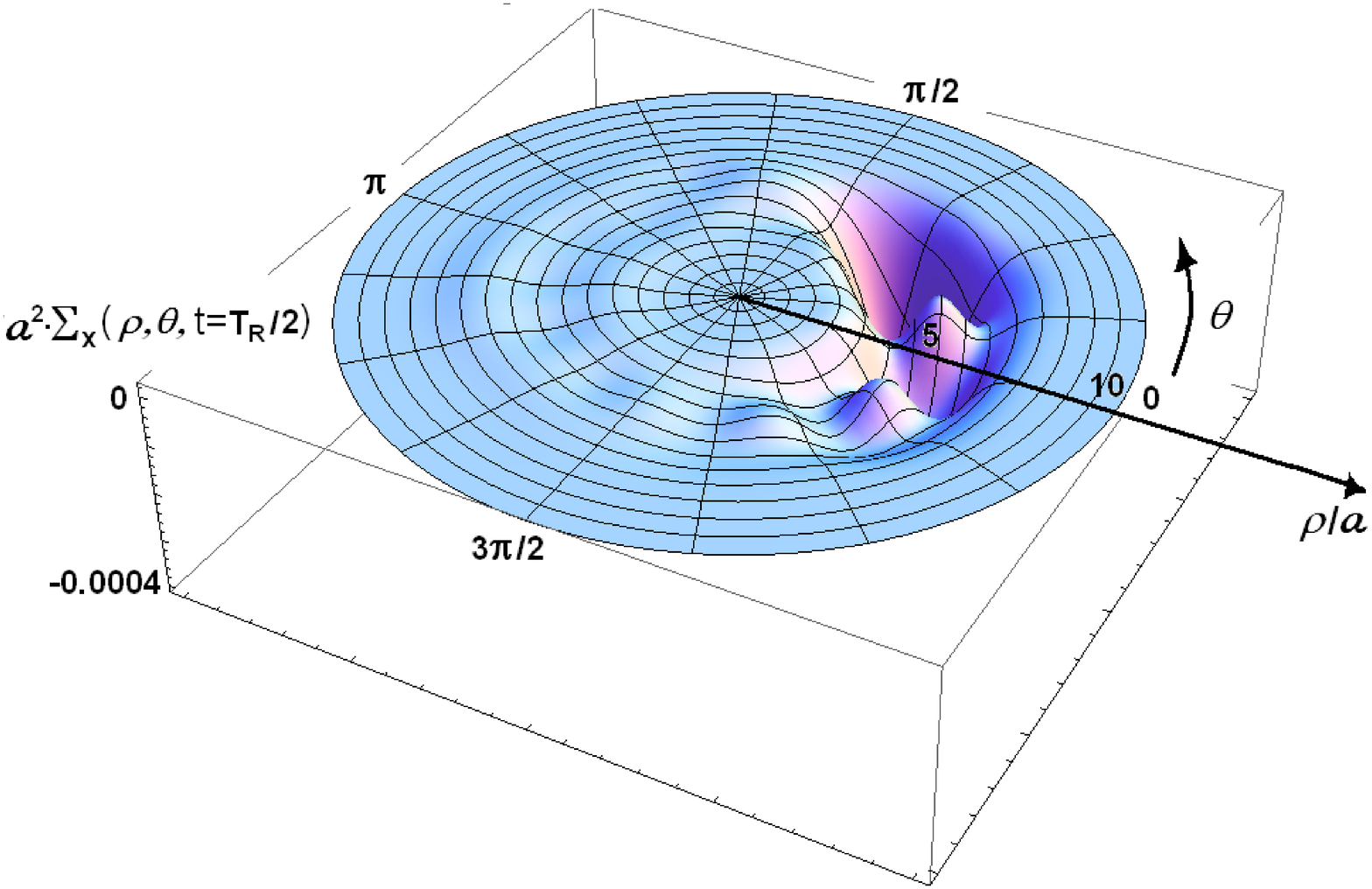}
(d)\includegraphics[width=0.4\textwidth]{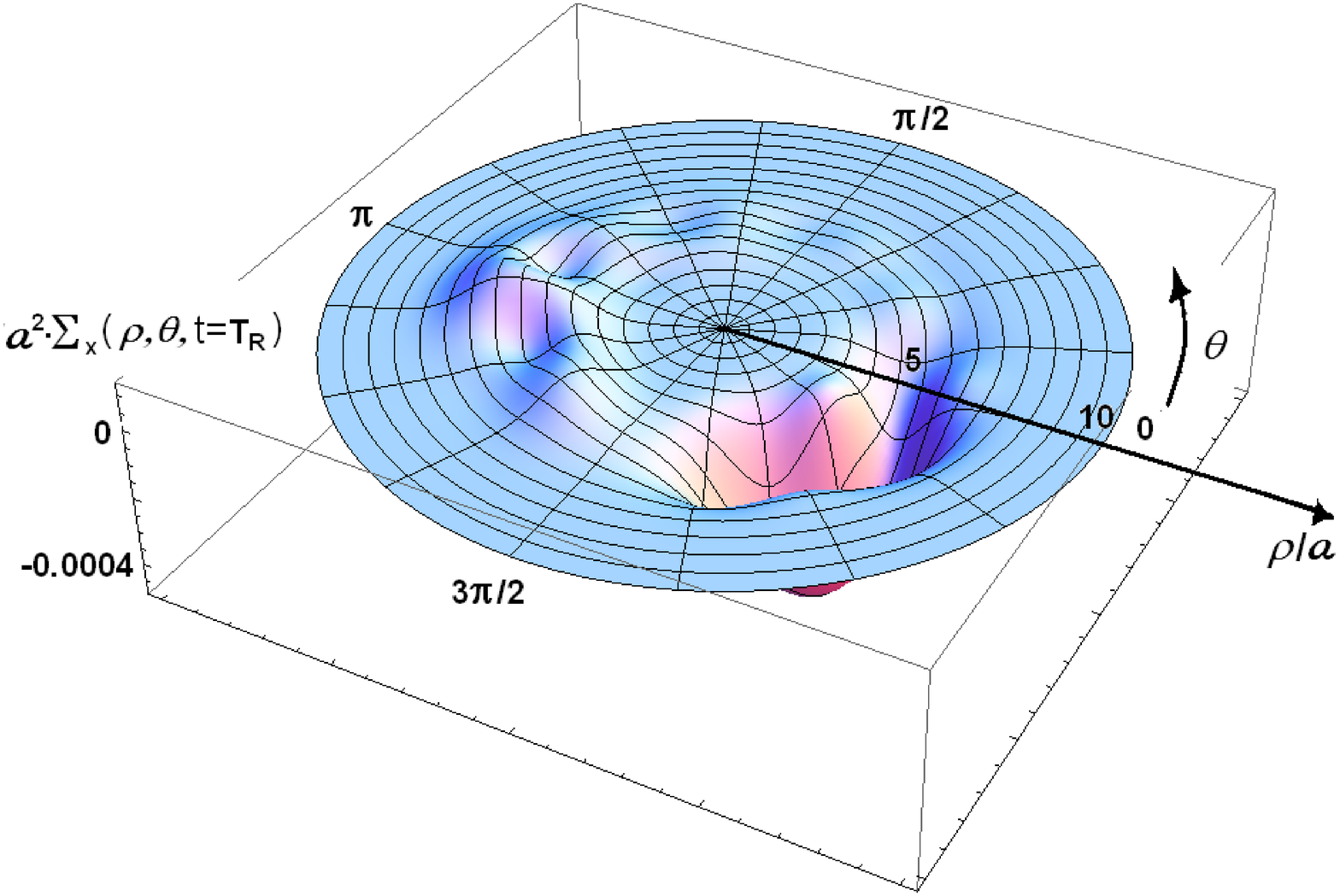}\caption{
(Color online) The distribution of spin density
$<\Sigma_x(\rho,\theta, t)>$ for the wave packet, Eq.(24), with
the parameters $\lambda/a=0.1$, $qa=5$ at times: (a) $t=0$, (b)
$t\approx T_R/4$, (c) $t\approx T_R/2$, (d) $t\approx T_R$.}
\end{center}
\end{figure}

\begin{figure}
\begin{center}
(a)\includegraphics[width=0.4\textwidth]{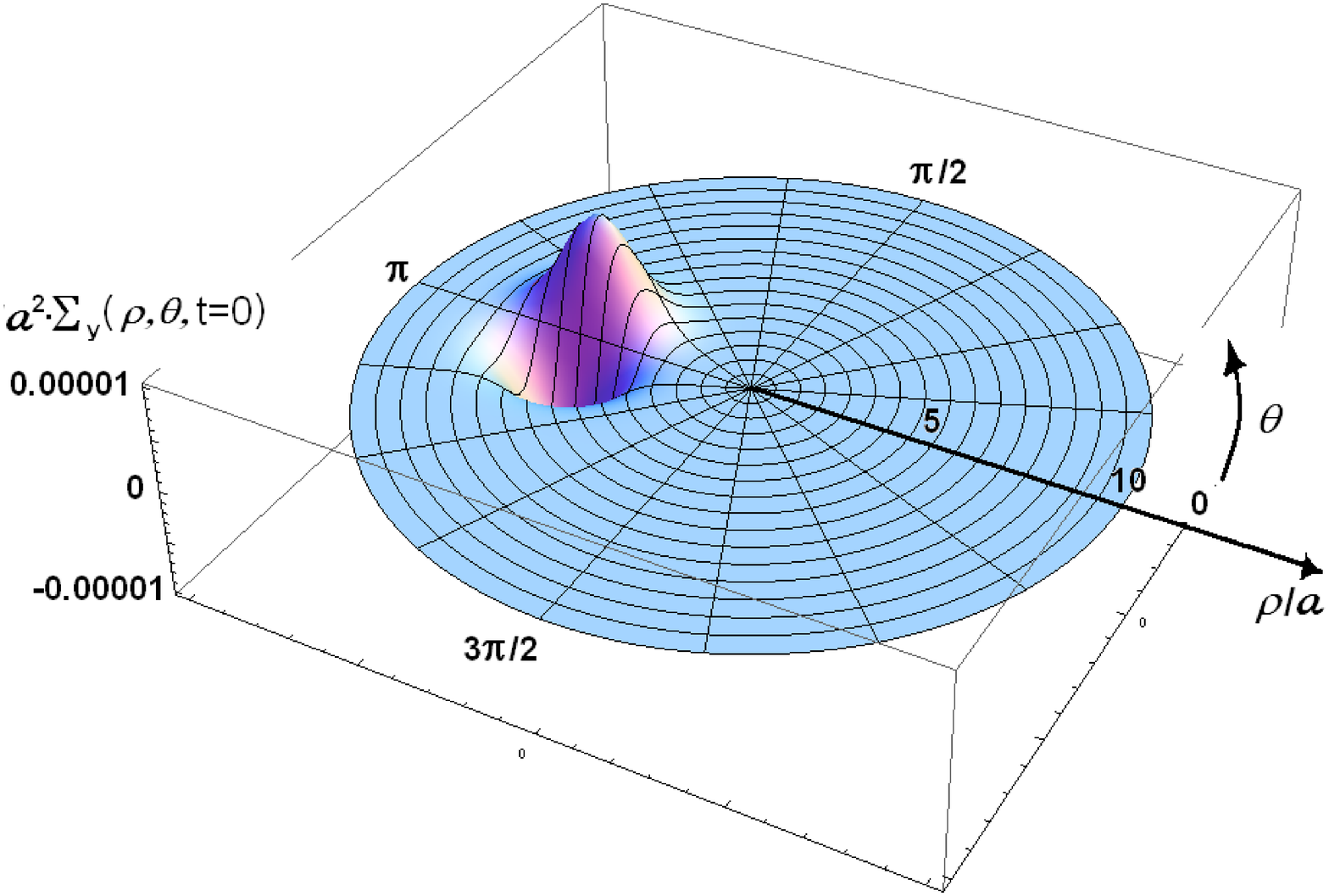}
(b)\includegraphics[width=0.4\textwidth]{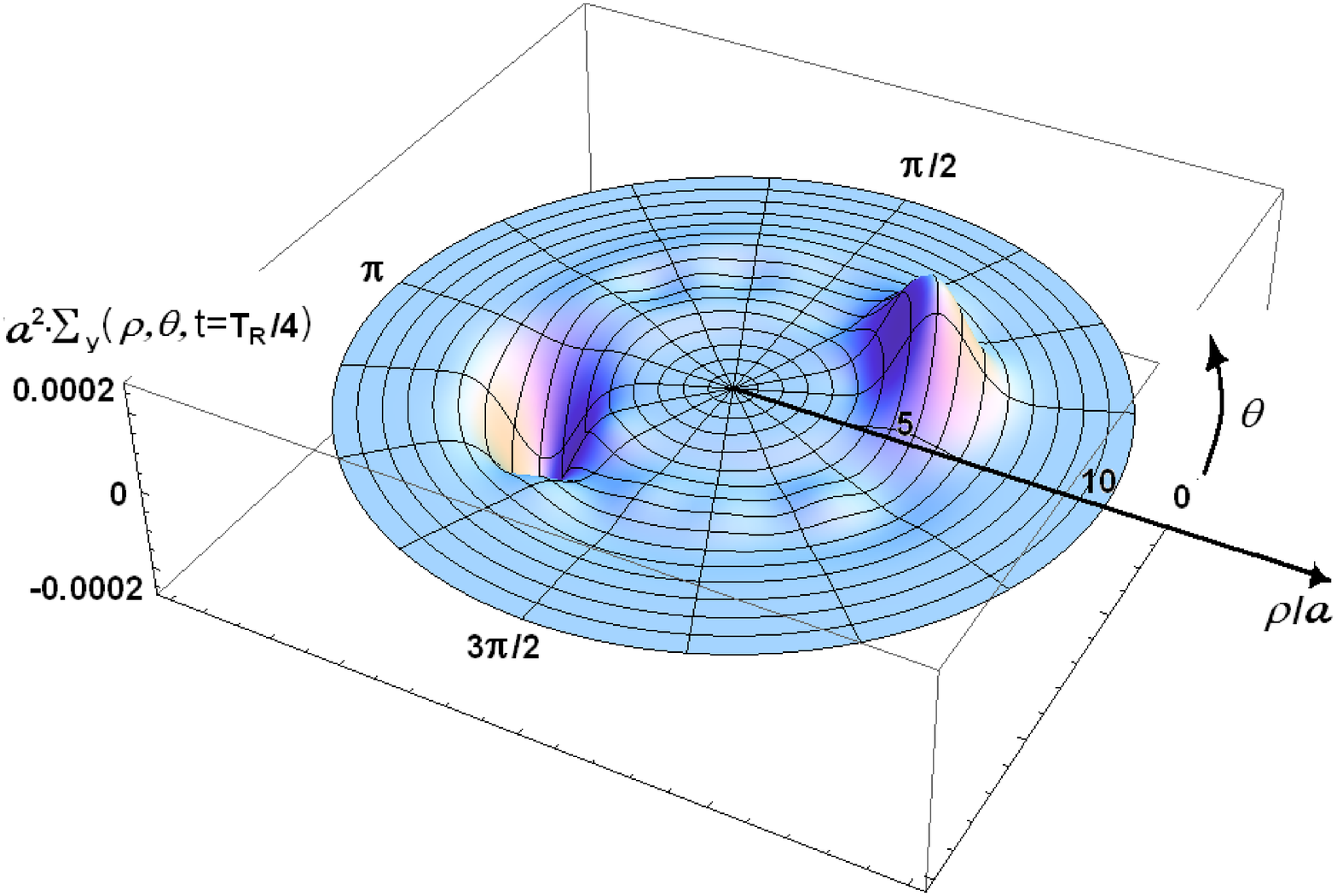}
\\
(c)\includegraphics[width=0.4\textwidth]{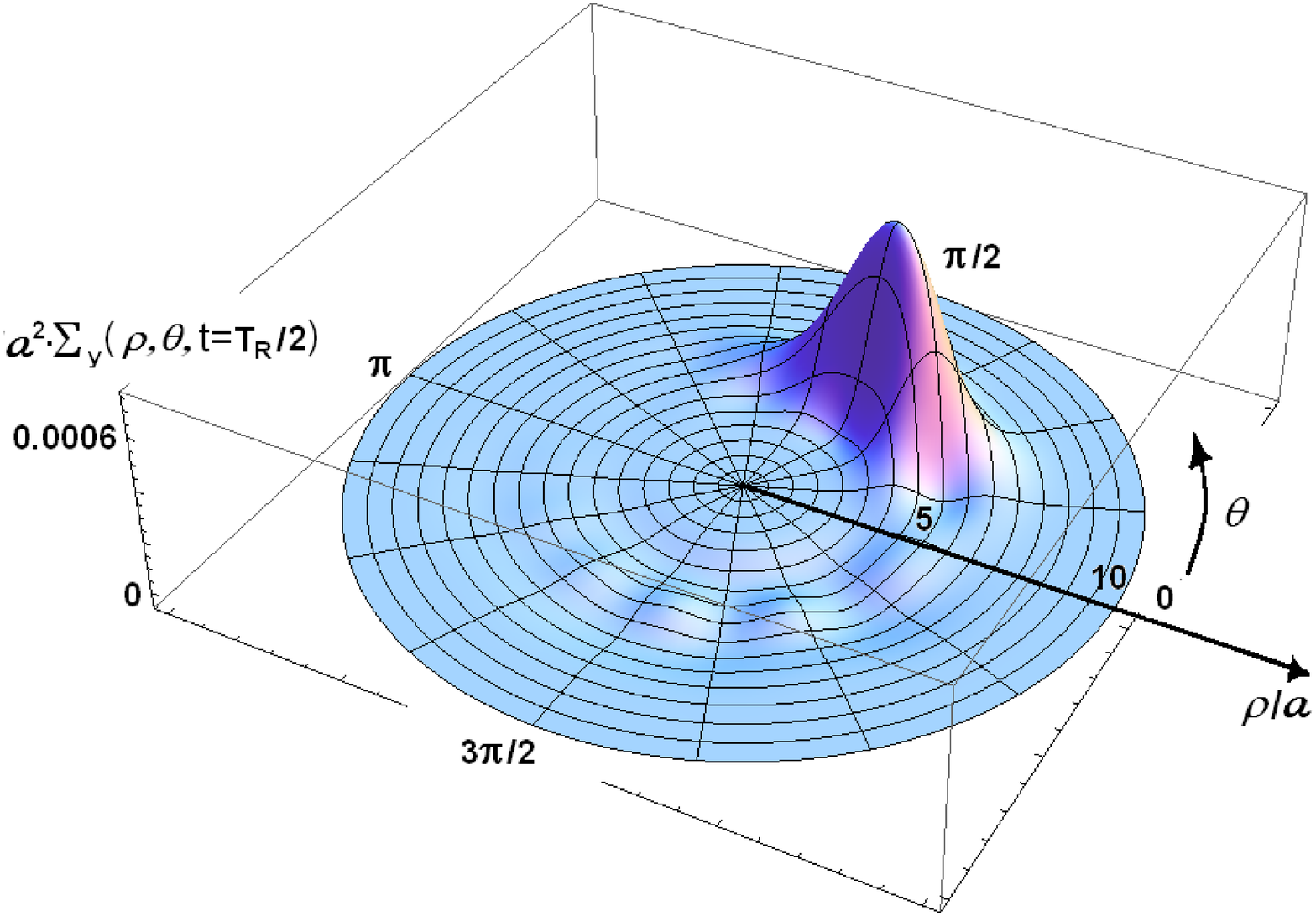}
(d)\includegraphics[width=0.4\textwidth]{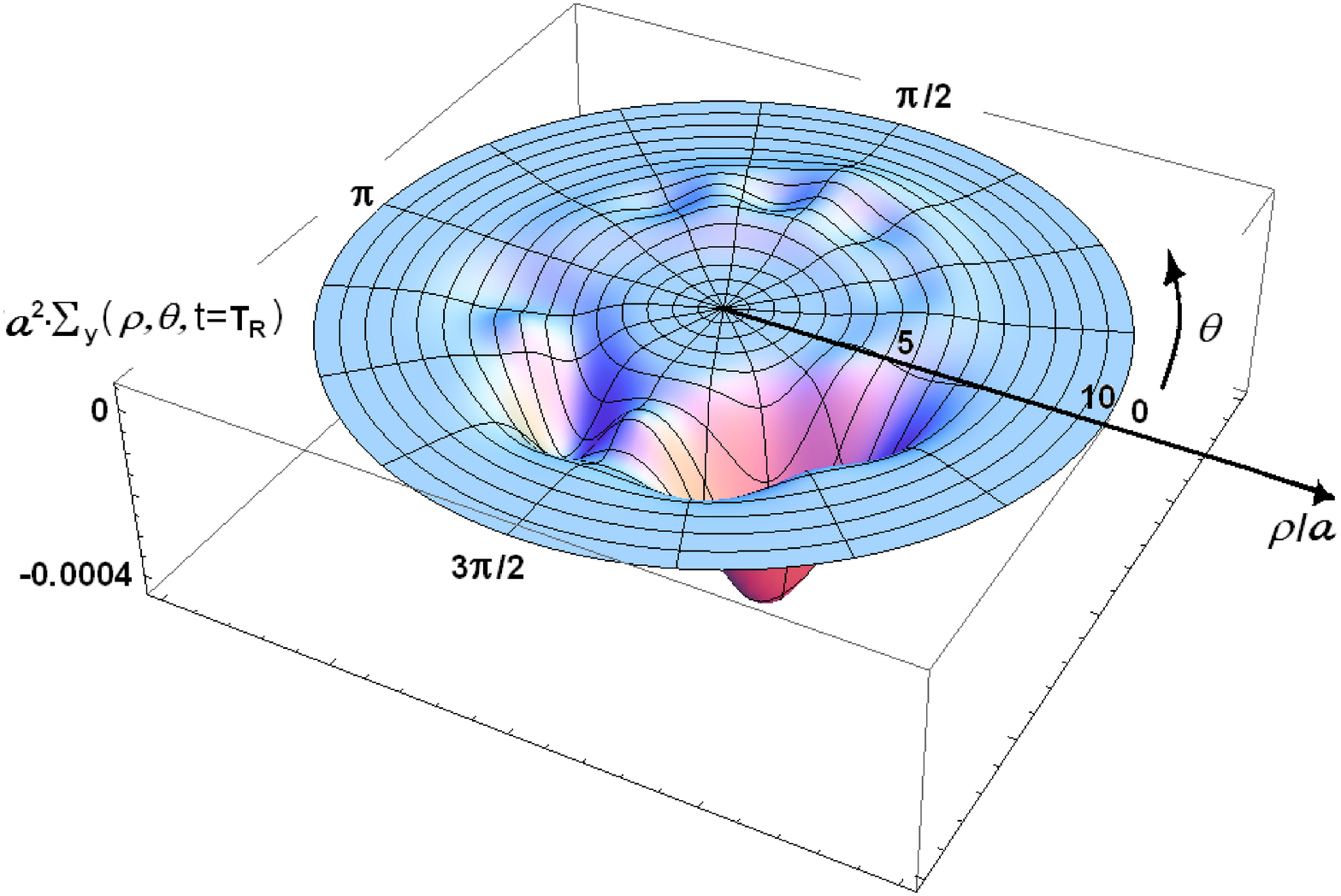}\caption{
(Color online) The distribution of spin density
$<\Sigma_y(\rho,\theta, t)>$ for the wave packet, Eq.(24), with
the parameters $\lambda/a=0.1$, $qa=5$ at times: (a) $t=0$, (b)
$t\approx T_R/4$, (c) $t\approx T_R/2$, (d) $t\approx T_R$.}
\end{center}
\end{figure}

Fig.4 and Fig.5 illustrate the distributions of
$<\Sigma_x(\rho,\theta , t)>$ and $<\Sigma_y(\rho,\theta , t)>$ at
the same moments as an Fig.2, i.e. $t=0,\, T_R/4,\, T_R/2,\, T_R$.
Just as for electron probability density at $t\lesssim T_R$ in a
main approximation the space-time evolution of the spin density is
determined by the wave function $\psi_{cl}({\bf r},t)$, Eq.(B4),
which corresponds to the "classical" motion of the electron. So
that let us define the functions ($\alpha=\beta$)
$$
\displaylines{
<\Sigma_x^{cl}(\rho,\theta ,t)>=\exp\Big(-\frac{\rho^2+(qa)^2+2\rho qa\cos(\theta+
2\pi t/T_c)}{2}\Big)\times\cr
\hfill\Big(\cos\frac{2\pi t}{T_c}-\frac{\lambda^2}{4a^2}(qa\rho)\cos\theta\Big)/
(2\pi a^2),\hfill\llap{(44)}\cr
}
$$
$$
\displaylines{
<\Sigma_y^{cl}(\rho,\theta ,t)>=\exp\Big(-\frac{\rho^2+(qa)^2+2\rho qa\cos(\theta+
2\pi t/T_c)}{2}\Big)\times\cr
\hfill\Big(\sin\frac{2\pi t}{T_c}+\frac{\lambda^2}{4a^2}(qa\rho)\sin\theta\Big)/
(2\pi a^2).\hfill\llap{(45)}\cr
}
$$
One can see from these expressions that at $t=0$ the $x$-component of spin density as a function
of an angular variable $\theta$ has a maximum at $\theta_0=\pi$ (Fig.4a). The maximum and
minimum values of the function $<\Sigma_y(\rho,\theta ,0)>$ are in the vicinity of
this point and differ in sign (see Fig.5a). At $t_2=T_R/4$, as was shown above,
we have two "classical" wave packets, Eq.(38), Fig.2c. Correspondingly,
$$
<\Sigma_{x,y}(\rho,\theta ,t_2)>=\frac{1}{2}(<\Sigma_{x,y}^{cl}(\rho,\theta ,t_2)>+
<\Sigma_{x,y}^{cl}(\rho,\theta ,t_2+T_{cl}/2)>),
\eqno{(46)}
$$
that is illustrated by Fig.4b and Fig.5b. Note that such simple summation in Eq.(46) is only
valid if the width of each of the wave packets is less than the cyclotron orbit length.

\section{Space-time dynamics of the mesoscopic wave packets formed by the positive and negative energy eigenstates}

In this Section we investigate the relativistic dynamics of the wave packet containing both
the positive and negative energy states with $\lambda_k=1$. Correspondingly we restrict
ourselves to the wave states with two nonzero components $\{\psi_1,\psi_4\}$. As was noted
in Section II the temporal dynamics of such states is governed by the part of Dirac
Hamiltonian which is identical to a Jaynes-Cummings interaction. Besides of
the general peculiarities of quantum dynamics of the system with the nonlinear energy
spectrum (collapse and revival phenomenon), the existence of two energy bands leads to
another effect -- {\it Zitterbewegung} (ZB). This phenomenon, generating a highly oscillatory
motion, is caused by the interference between positive and negative energy components
of the wave packet. We will consider here the effect of ZB on the evolution of the electron
velocity as well as on the spin polarization.

Let at the time $t=0$ the initial wave packet is given by
$$
\psi({\bf r},0)=\psi_c({\bf r})\pmatrix{|\uparrow>\cr 0\cr},
\eqno{(47)}
$$
where
$$
\psi_c({\bf r})=\frac{1}{a\sqrt{2\pi}}\exp\Big(-\frac{x^2+y^2}{4a^2}+
iqx+\frac{ixy}{2a^2}\Big)
\eqno{(48)}
$$
is the wave function of coherent state, which can be written in the form
$$
\psi_c({\bf r})=\int dp\, \varphi_p(x)g(p)\sum_{n=0}c_{n+1}\phi_n(y-y_c).
\eqno{(49)}
$$
The coefficients $g(p)$ and $c_n$ in this expression are determined by Eqs.(22)
and (23). Using Eqs.(20), (47) and (49), one can obtain
$$
\displaylines{
\psi({\bf r},\tau)=\int dp\, \varphi_p(x)g(p)\times\cr
\hfill\sum_{n=1}\bigg[d_n\pmatrix{d_n\phi_{n-1}(y-y_c)|\uparrow>\cr
-b_n\phi_n(y-y_c)|\downarrow>\cr}\exp(-i\varphi_n\tau)+
b_n\pmatrix{b_n\phi_{n-1}(y-y_c)|\uparrow>\cr
d_n\phi_n(y-y_c)|\downarrow>\cr}\exp(i\varphi_n\tau)\bigg].\hfill\llap{(50)}\cr}
$$
Performing integration over $p$ (see Appendix B, Eq.(B3)), we get
finally
$$
\psi({\bf r},\tau)=\Psi_1({\bf r},\tau)\pmatrix{|\uparrow>\cr 0\cr}+
\Psi_2({\bf r},\tau)\pmatrix{0\cr |\downarrow>\cr},
\eqno{(51)}
$$
where
$$
\Psi_1({\bf r},\tau)=\psi_c({\bf r})\exp\Big(-\frac{q}{2}(qa^2-y+ix)\Big)
\sum_{n=1}\frac{(q(qa^2-y+ix))^{n-1}}{2^{n-1}(n-1)!}\Big(
\cos(\varphi_n\tau)-\frac{i\sin(\varphi_n\tau)}{\varphi_n}\Big),
\eqno{(52)}
$$
$$
\Psi_2({\bf r},\tau)=-\psi_c({\bf r})\exp\Big(-\frac{q}{2}(qa^2-y+ix)\Big)
\sum_{n=1}\frac{(q(qa^2-y+ix))^{n-1}}{2^{n-1}(n-1)!}
\frac{i\lambda\sin(\varphi_n\tau)}{a^2\varphi_n}(qa^2-y+ix).
\eqno{(53)}
$$

As above, first of all we calculate the average velocity of the packet center
$$
\overline{v_i}(t)=c\int \psi^{\dagger}\hat\sigma_i\psi\, d{\bf r}.
\eqno{(54)}
$$
Substituting the expressions for the components of the wave function Eqs.(52),
and (53) into Eq.(54) after the integration over the $x$, $y$ coordinates we have
$$
\overline{v_x}(\tau)=c\frac{\lambda}{a}\exp\Big(-(qa)^2/2\Big)
\sum_{n=1}\frac{(qa)^{2n-1}}{2^{n-1}(n-1)!\varphi_n\varphi_{n+1}}
\Big(\cos((\varphi_{n+1}-\varphi_n)\tau)-\cos((\varphi_{n+1}+\varphi_n)\tau)\Big),
\eqno{(55a)}
$$
$$
\overline{v_y}(\tau)=c\frac{\lambda}{a}\exp\Big(-(qa)^2/2\Big)
\sum_{n=1}\frac{(qa)^{2n-1}}{2^{n-1}(n-1)!\varphi_n}
\Big(\sin((\varphi_{n+1}-\varphi_n)\tau)-\sin((\varphi_{n+1}+\varphi_n)\tau)\Big).
\eqno{(55b)}
$$

\begin{figure}
  \centering
  \includegraphics[width=150mm]{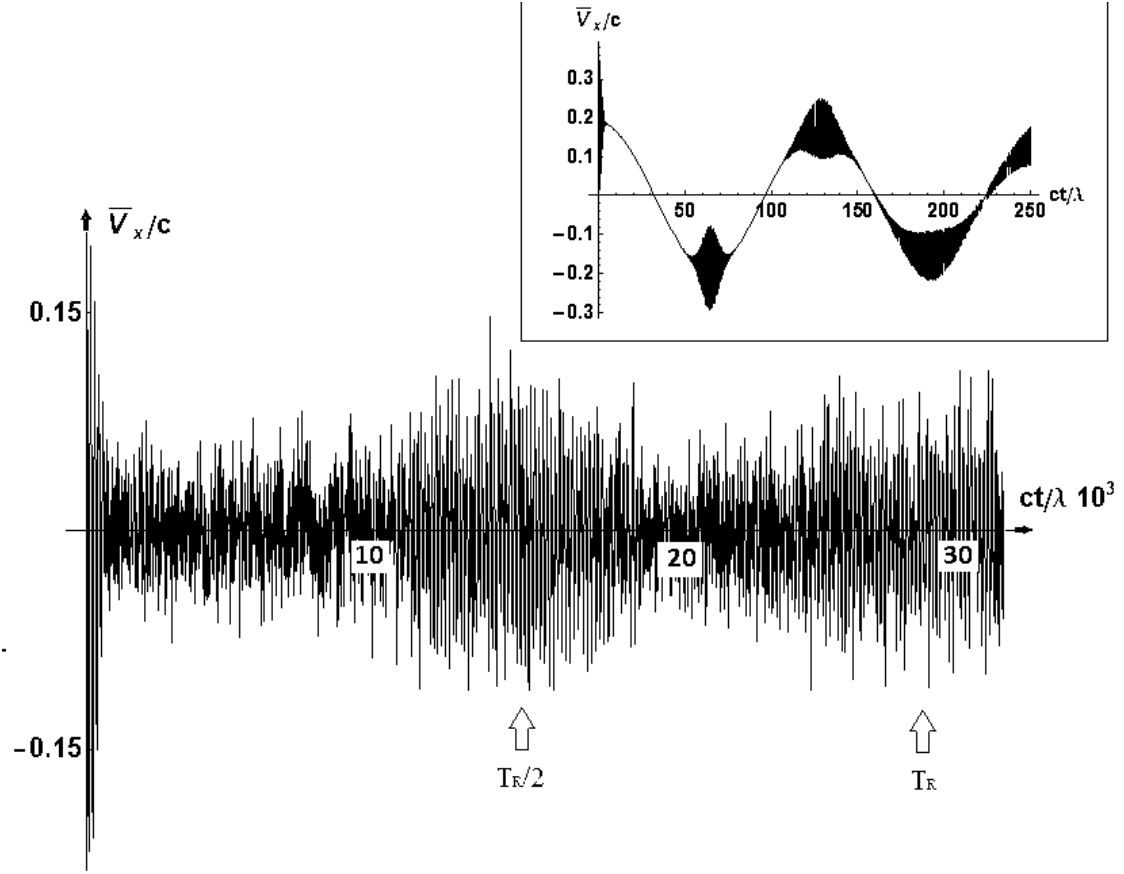}
  \caption{The dependence of the average velocity $\overline{v_x}(t)$ versus time
for the initial wave packet, Eq.(47), with the parameters
$\lambda/a=0.5$, $qa=10$. The oscillations of the average velocity
$\overline{v_x}(t)$ at times $t\geq T_{cl}$ are shown in the
insert.}
\end{figure}

Note that these equations depend on not only on the energy
difference $\varphi_{n+1}-\varphi_n$ as in Eq.(25), but on the sum
$\varphi_{n+1}+\varphi_n$ also. The last dependence leads to the
high frequency of the ZB oscillations. The dependence of the
average velocity $\overline{v_x}(\tau)$ on time for the packets
parameters $\lambda/a=0.5,\, qa=10$ (that corresponds to the
dominant value of $n=n_0\simeq 50$) is shown in Fig.6. Like the
case of the wave packet discussed in Section II, the behavior of
this function is accompanied by the collapse and revival
phenomenon. For the parameters of our wave packet the
corresponding periods of cyclotron motion and revivals are
$T_{cl}\approx 126\lambda/c$ and $T_R\approx 26660\lambda/c$. In
addition, the insert in Fig.6 clearly demonstrates the ZB
oscillations with peculiar frequency
$\omega_{ZB}=2\varphi_{n_0}c/\lambda=10.2 c/\lambda\simeq
7.85\cdot 10^{21} \, s^{-1}$. Besides, these oscillations occur in
short intervals near time moments $t_k=kT_{cl}/2\,
(k=1,2,\ldots)$, that is connected with the feature of the
space-time evolution of the wave packet probability density. At
$t>0$ the initial wave packet (Fig.7a) splits into two parts
(Fig.7b) with the amplitudes which are slightly different. These
sub-packets containing the states of one sign of the  energy
rotate with cyclotron frequency in the opposite directions and
meet each other every one-half of the cyclotron period (see
Fig.7c, 7d). The duration of ZB-oscillations is determined by the
relation of the packets size to their relative velocity. Note that
in the moment of time $t=T_{cl}/4$ the spin parts of the sub-wave
packets with $E_n>0$ and $E_n<0$ differ slightly. Thus, the
spatial part of the full wave function (50) is a linear
superposition of the mesoscopic states (see Appendix C). At $t\sim
t_k$ the parts with positive and negative energy have significant
overlap in the coordinate space. This is a necessary condition for
the existence of {\it Zitterbewegung} (see insert in Fig.6).

At times $t\sim T_{cl}$ the effect of the quadratic term in the
Taylor expansion of the energy, Eq.(A2), is insignificant. For
times much greater than the cyclotron period the dephasing appears
in individual terms in Eq.(50) due to the term $(n-n_0)^2$, that
leads to the collapse of the sub-wave packets. At intermediate
times $T_{cl}\ll t_n\leq T_R$, where $t_n\approx mT_R/n$ ($m/n$ is
irreducible fraction), the fractional revival of each sub-packets
occurs. As a result, each sub-packet is decomposed into
$N=n(3-(-1^n))/4$ packets-fractions. Particularly, at $t=T_R$ each
of the two sub-packets (with positive or negative energy) restores
at various points on the cyclotron orbit, that makes it impossible
reshaping of the initial wave packet entirely. At time $T_R/4$ we
should see four packets-fractions. However, this result is not
consistent with the distribution of the probability density shown
in Fig.8a due to the significant influence of the term $\sim
(n-n_0)^3$ in the series (A2) for times $t\leq T_R$. This
statement is illustrated in Fig.8b, which shows at time $t=T_R/4$
the probability density, taking into account the first three terms
in the expansion (A2) only.

Consider now the average value of the spin operator. Since the initial state
of the wave packet, Eq.(47), belongs to one of the invariant subspaces (with $\lambda_k=1$),
the average spin projections $\overline{S_x}=\overline{S_y}=0$. Using the definition
$\overline{S_z}=(\hbar/2)\hat\sigma_z$ and Eqs.(51), (52), and (53) we obtain

\begin{figure}
\begin{center}
(a)\includegraphics[width=0.4\textwidth]{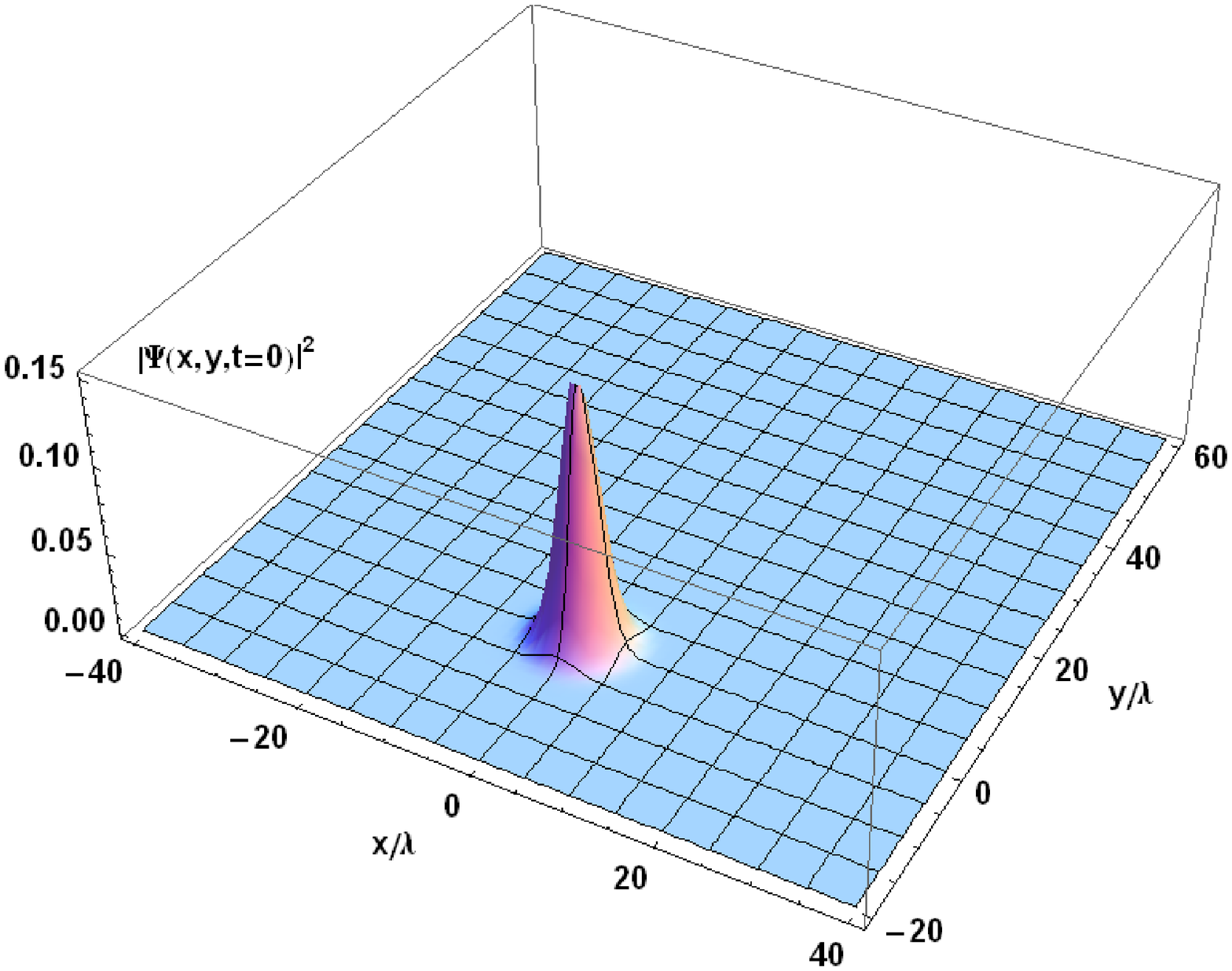}
(b)\includegraphics[width=0.4\textwidth]{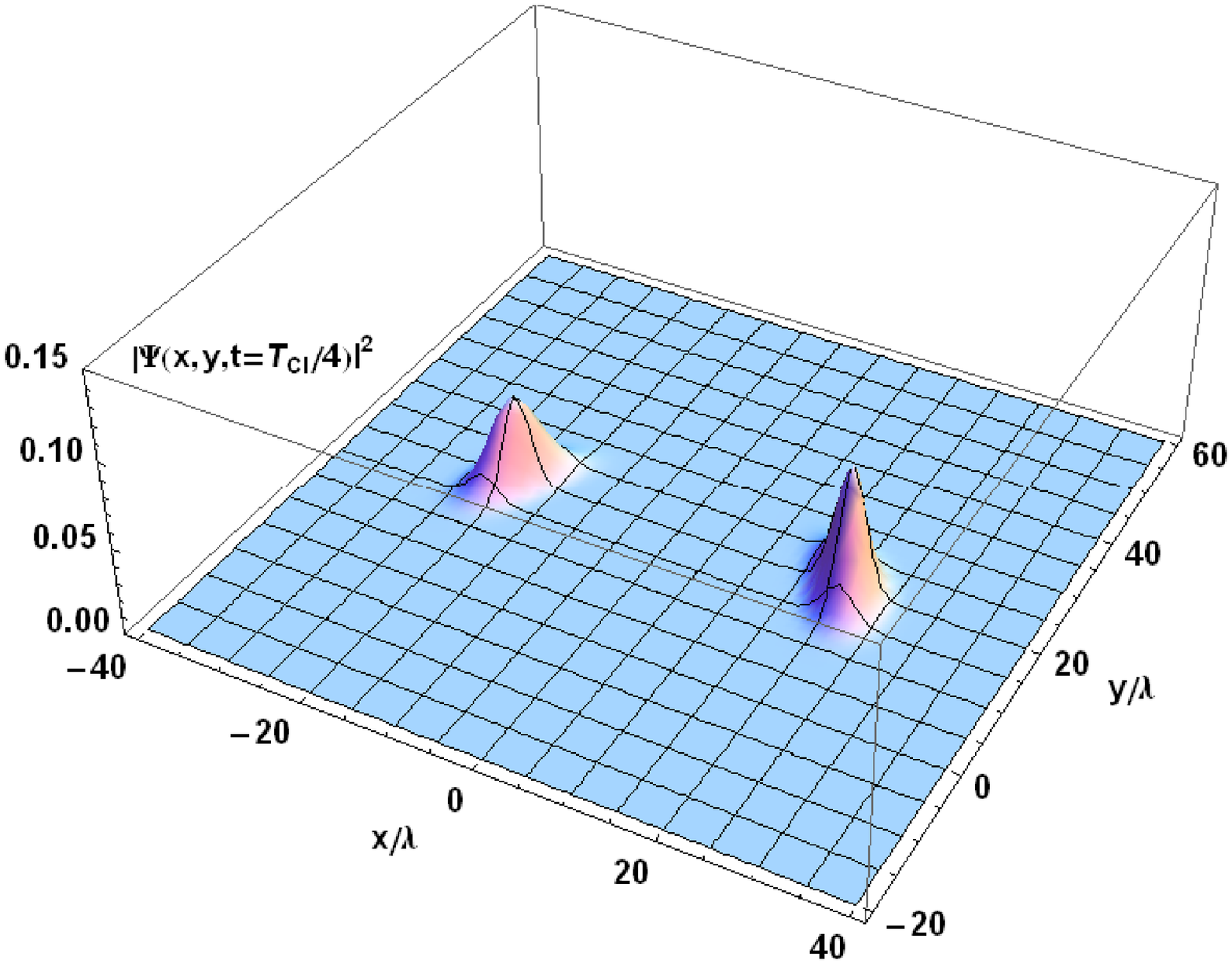}
\\
(c)\includegraphics[width=0.4\textwidth]{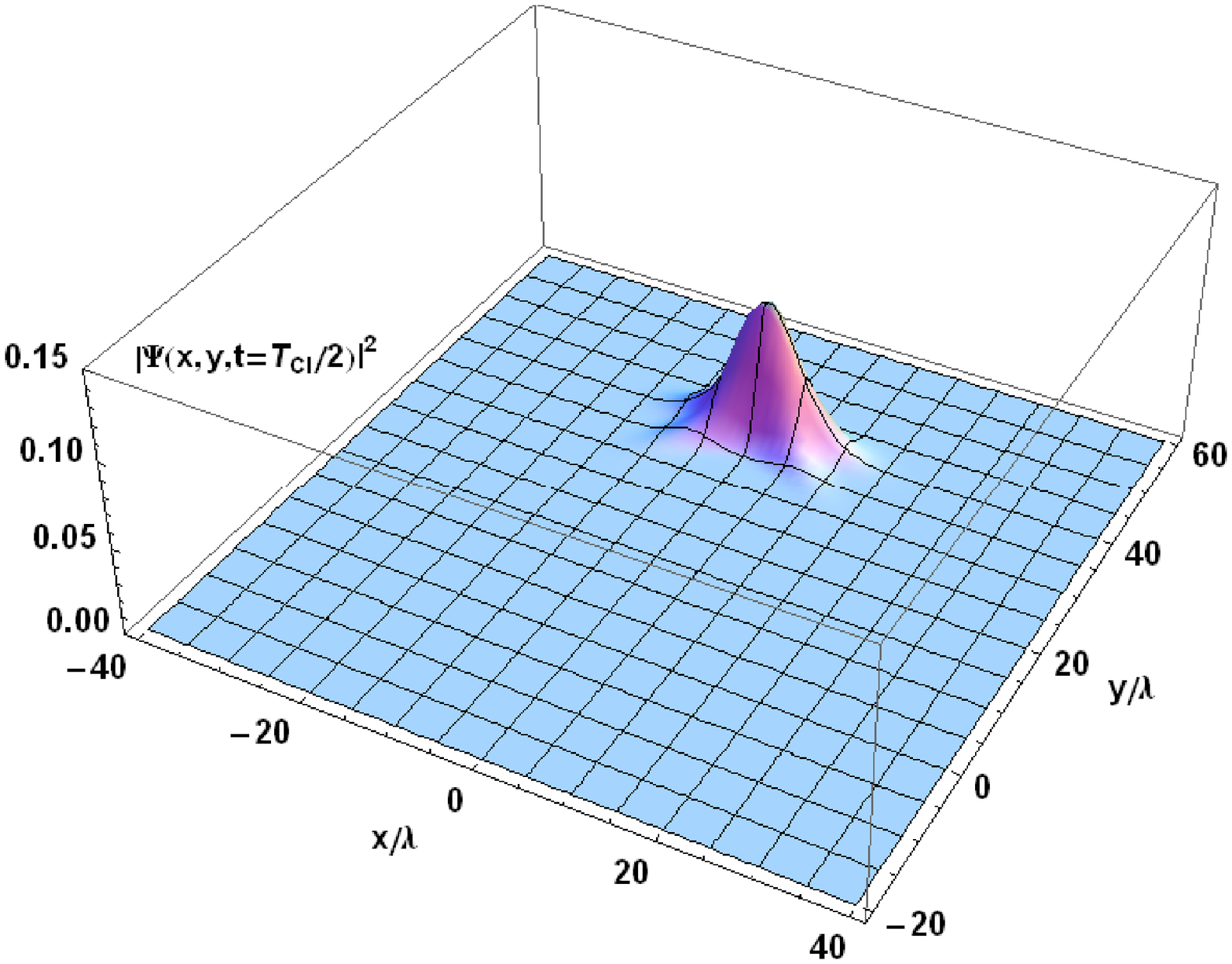}
(d)\includegraphics[width=0.4\textwidth]{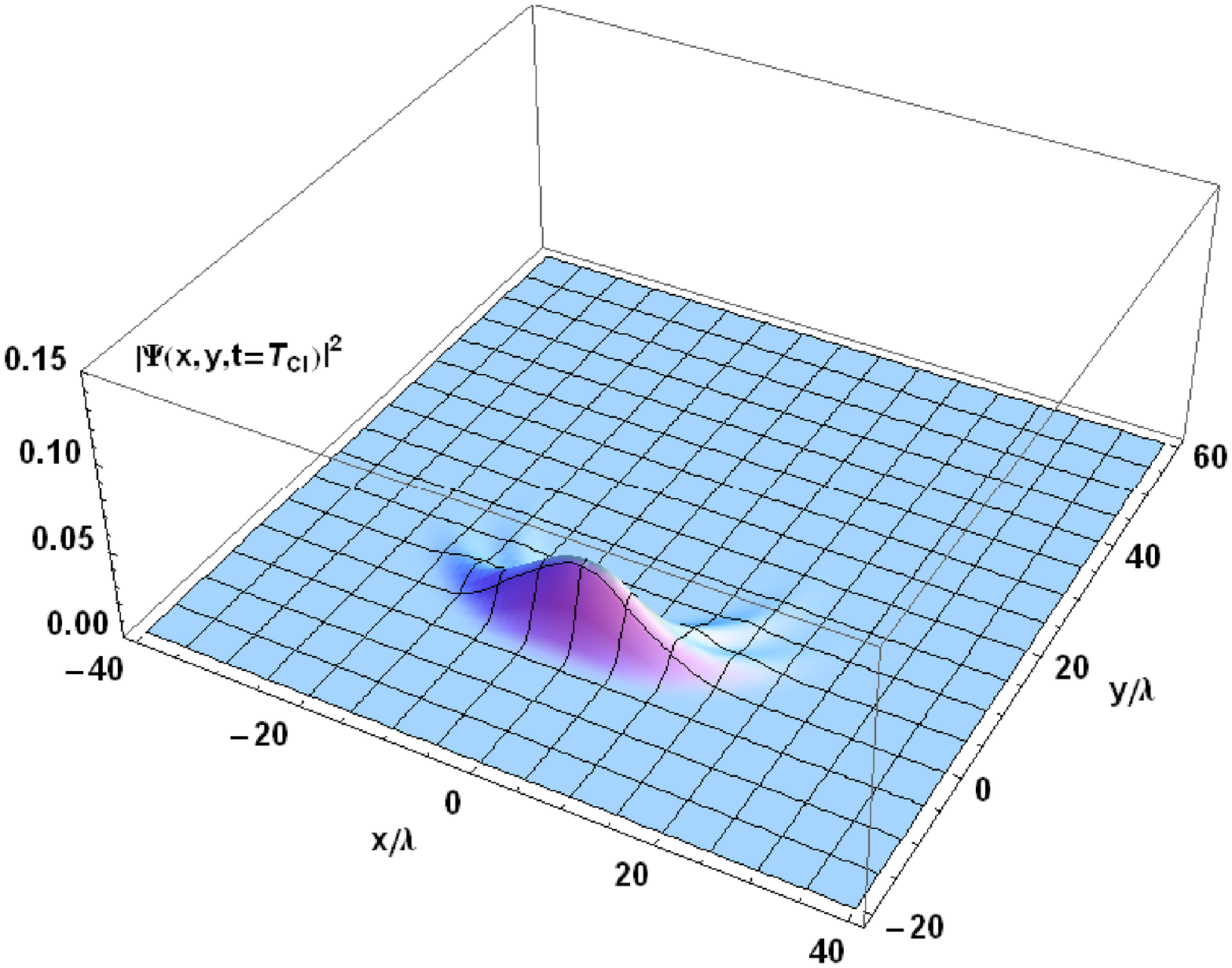}\caption{
(Color online) The distribution of the electron probability
density for initial wave packet, Eq.(47), with the parameters
$\lambda/a=0.5$, $qa=10$: (a) $t=0$, (b) $t\approx T_{cl}/4$, (c)
$t\approx T_{cl}/2$, (d) $t\approx T_{cl}$.}
\end{center}
\end{figure}

$$
\overline{S_z}(\tau)=\frac{\hbar}{2}\sum_{n=1}|c_n|^2\bigg(\frac{1+2n(\lambda/a)^2
\cos(2\varphi_n\tau)}{\varphi_n^2}\bigg),
\eqno{(56)}
$$
where the coefficients $c_n$ are determined by Eq.(23).
Corresponding dependence is shown in Fig.9. It is clearly seen
that the average value of the $z$-component of spin oscillates
with the ZB-frequency $\omega_{ZB}\approx 2\varphi_{n_0}c/\lambda$
and is accompanied by the phenomenon of collapse and revival. In
this case unlike the previous definition, Eq.(30), the
corresponding revival time is determined by the cyclotron period
$T_{rev}=T_{cl}/2$.\cite{{Berm,Gea}} In intervals between the
ZB-oscillations the average value of $\overline{S_z}(t)$ is not
equal to zero. As mentioned above, the space-time evolution of the
original wave packet (47) is described by the part of the Dirac
Hamiltonian, Eq.(1) that is identical to the Jaynes-Cummings model
in quantum optics.\cite{JC} So one can conclude that Eq.(56) is
similar to the well-known expression for the population of the
ground state of two-level atom exposed to the action of the
classical electromagnetic field\cite{A1}.

\begin{figure}
\begin{center}
(a)\includegraphics[width=0.4\textwidth]{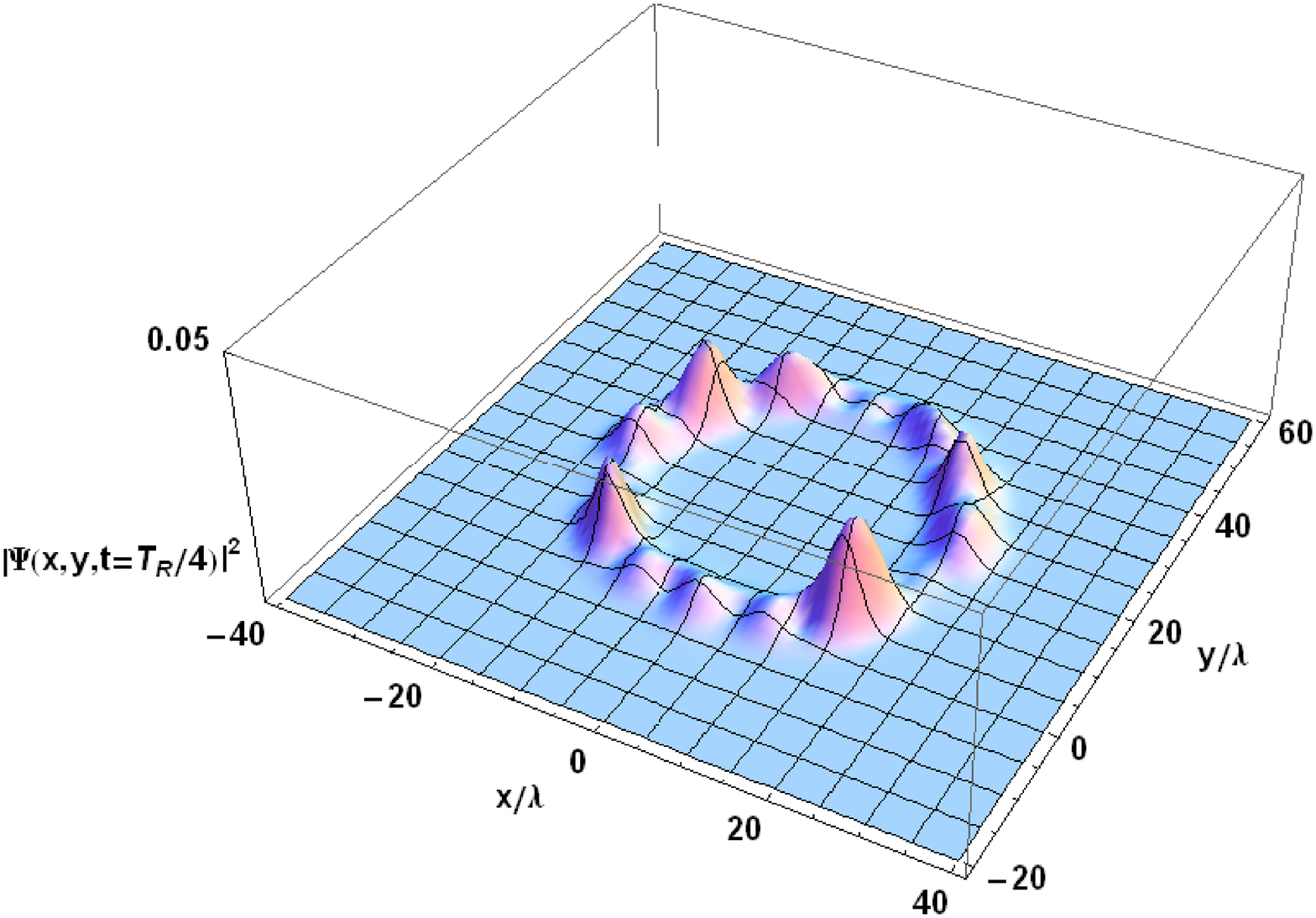}
(b)\includegraphics[width=0.4\textwidth]{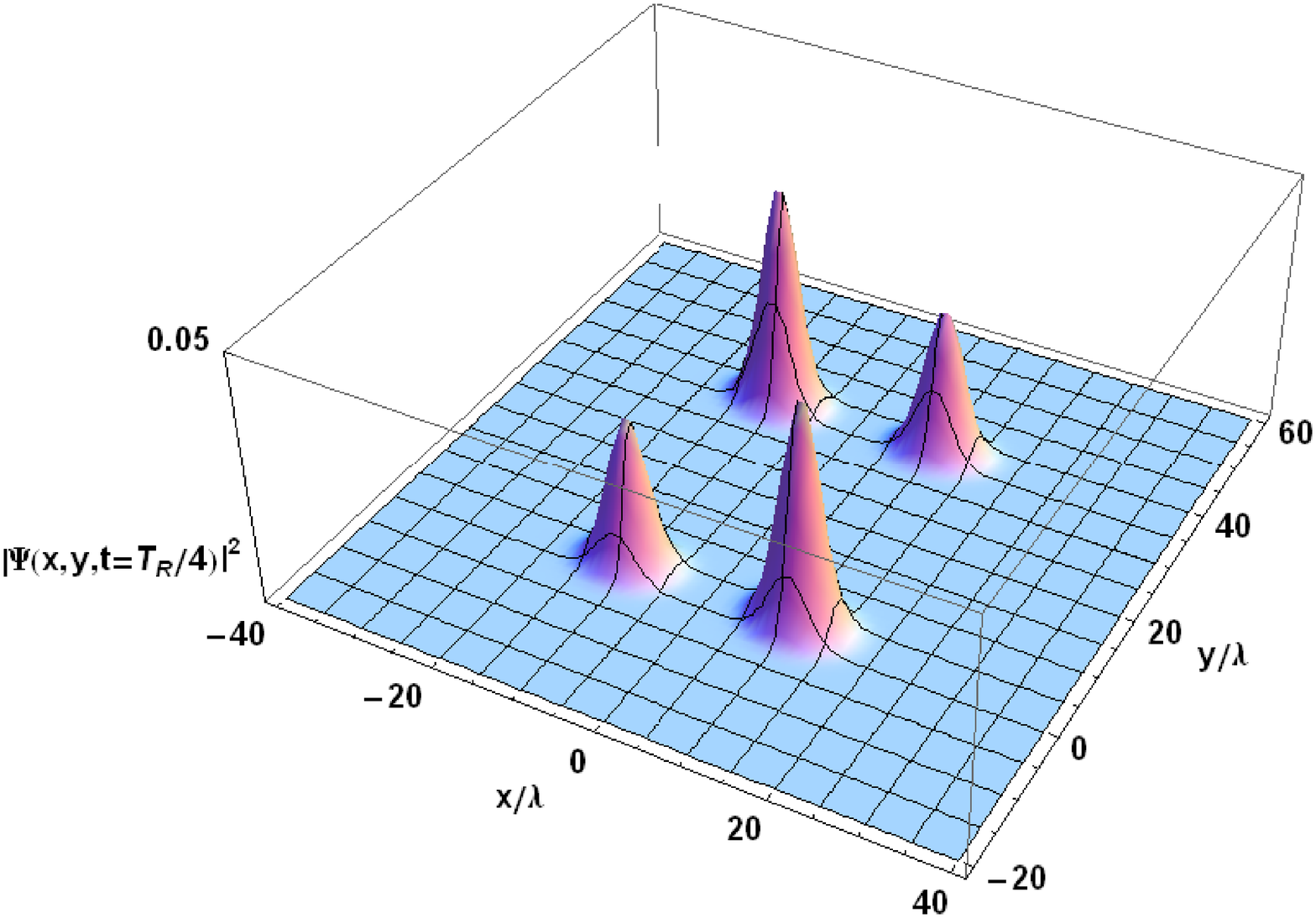} \caption{
(Color online) Electron probability density for initial wave
packet, Eq.(47), at $qa=10$, $\lambda/a=0.5$, $t\approx T_R/4$:
(a) for real energy spectrum, (b) for the spectrum obtained by
taking into account the first three terms in the expansion (A2)
only.}
\end{center}
\end{figure}

\begin{figure}
  \centering
  \includegraphics[width=150mm]{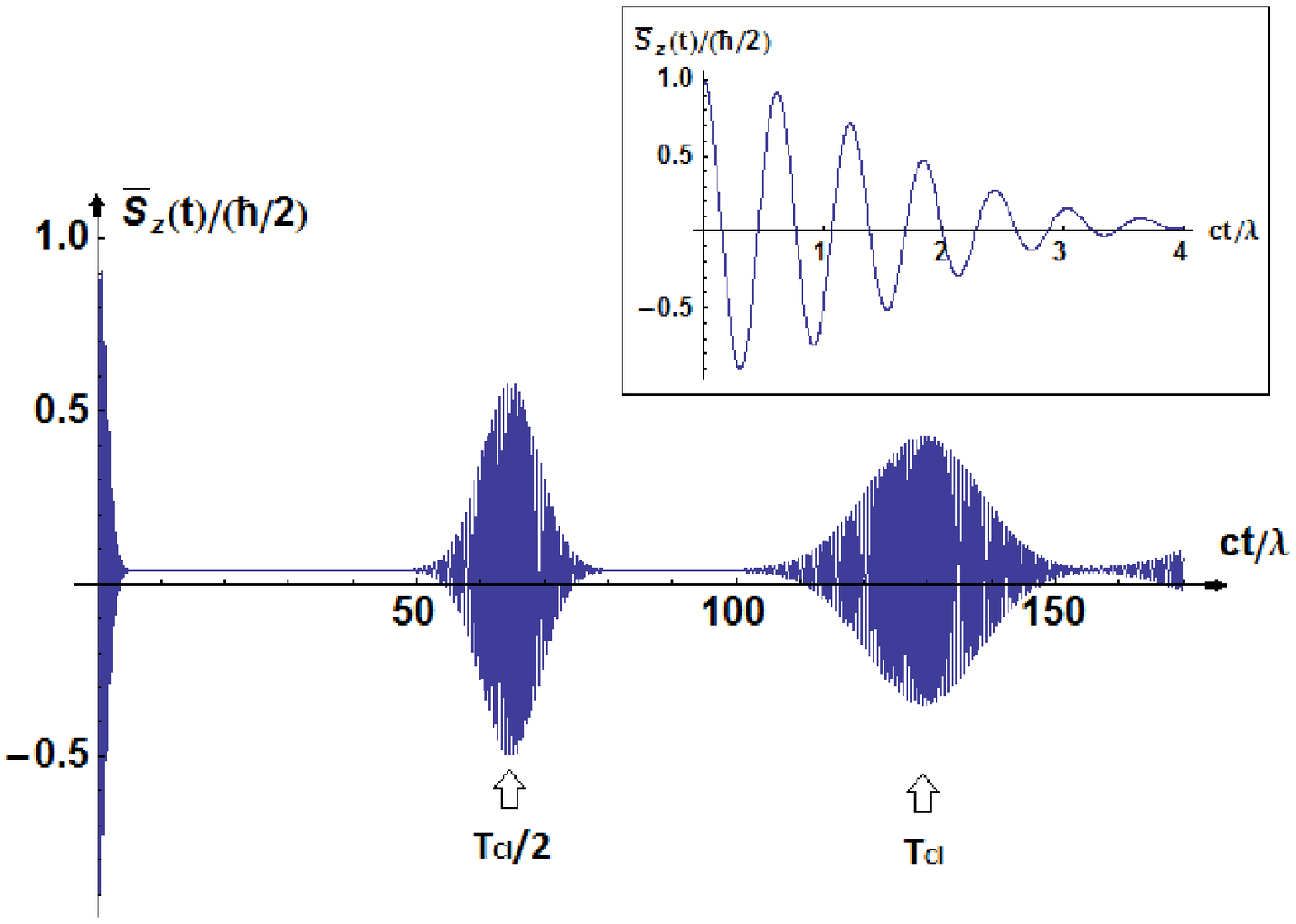}
  \caption{(Color online) The time dependence of average spin $\overline{S_z}$ (in units of $\hbar/2$)
for initial wave packet, Eq.(47), for $qa=10$, $\lambda/a=0.5$.
The oscillations of the average spin at short times $t\ll T_{cl}$
are shown in the insert.}
\end{figure}

\section{Summary}

In this work we have analyzed the cyclotron dynamics of
relativistic wave packets, moving in external constant magnetic
field over cyclotron orbit. The components of electron wave
function were calculated (see Eqs. (31), (52), and (53)), as well
as, average velocities of the packet $\overline{v_x}(t)$. It was
shown that for the case when the wave packet consists of positive
energy states only the effects of collapse and revival of electron
and spin densities can be observed.

On short-time $t\leq T_{cl}$ (where $T_{cl}$ is the cyclotron
period) the coherent wave packet rotates with the classical
frequency over the cyclotron orbit. We analyzed the complex
evolution of spin densities and average spin components. The
precession of spin densities one can see in Fig.4, Fig.5. We
visualized $N$ packets-fractions which form at times of fractional
revivals $t=mT_R/n$ ($N=n(3-(-1)^n)/4$, as was shown by Averbukh
and Perel'man\cite{A1}). In our opinion, the dynamics of such
structures can be studied experimentally by investigating the
nature of electromagnetic multipole radiation and absorption. It
was found that the effects of fractional revivals can not be
observed via time-dependence of average velocity.

We compared the dynamics of wave packet described above with the
behavior of the wave packet containing the states with both the
positive and negative energies (Sec.IV). In the last case the
splitting of the initial wave packet takes place at $t<T_{cl}$.
These two sub-packets rotate with the cyclotron frequency in
opposite directions and meet each other every one-half of the
cyclotron period that results in the high-frequency ZB
oscillations. It was shown (see Appendix C) that of time
$t=T_{rev}/2=T_{cl}/4$ the electron spin and orbital degrees of
freedom become disentangled. In this case spin states play the
role of the internal states of the atom that has or has not
radioactively decayed and the orbital states correspond to the cat
in the original Schr\"odinger model\cite{Sh}. At $t=T_R$ each of
the two sub-packets restores at different points of the cyclotron
orbit. So that the full reshaping of initial wave packet is
impossible unlike the case of wave packet including the positive
energy states only.

\section{Acknowledgments}

The authors are grateful to M.A. Martin-Delgado, who drew their attention to the problem
of "Dirac cat", discussed in Ref.\cite{Berm}. This work was supported by a program of the Russian
Ministry of Education and Science "The development of scientific potential of higher
education"  (Project No. 2.1.1/10910), and Russian Foundation for Basic Research
(Grant No. 11-02-00960a).

\section*{Appendix A}

In this Appendix we find the approximate expression for the average electron velocity for the packet
parameters $qa\gg 1$ and $\lambda/a\ll 1$ (we consider $qa\simeq 5,\, \lambda/a\simeq 0.1$).
Corresponding wave function (20) represents the quantum superposition of the states
sharply peaked around large enough central value $n=n_0$. In fact, for $(qa)^2/2\gg 1$ the
Poisson distribution $|c_n|^2$ (see Eq.(23)) can be replaced by Gaussian one
$$
e^{-(qa)^2/2}\frac{\Big(\frac{(qa)^2}{2}\Big)^n}{n!}\approx\frac{1}{\sqrt{2\pi n_0}}
exp\bigg(-\frac{(n-n_0)^2}{2n_0}\bigg),
\eqno{(A1)}
$$
where $n_0\simeq (qa)^2/2\simeq 12$ and dispersion $\bigtriangleup
n\simeq \sqrt n_0$. So, we expand the dimensionless energy
spectrum $\varphi_n$ as a power series in $(n-n_0)$, restricting
ourselves to the quadratic terms\cite{{A1,Rom}}
$$
\varphi_n=\varphi_{n_0}+\varphi '_{n_0}(n-n_0)+\frac{\varphi ''_{n_0}}{2}(n-n_0)^2.
\eqno{(A2)}
$$
For the given values of the packet parameters $\varphi_n$ is close to $1$ and the
coefficients before the exponents in Eq.(25) can be estimated, in good approximation, as
$$
\sqrt{\frac{\varphi_{n+1}-1}{2(n+1)\varphi_{n+1}}}\simeq\frac{\lambda/a}{\sqrt{2}},
\; \sqrt{\frac{\varphi_n+1}{\varphi_n}}\simeq\sqrt{2},
\eqno{(A3)}
$$
that allows us to carry out the sum over $n$ in Eq.(25). The result is
$$
\displaylines{
\overline{v_x}(\tau)=\frac{\hbar q}{m}\exp\bigg(-\Big(qa\sin
\frac{\varphi ''_{n_0}\tau}{2}\Big)^2\bigg)\times\cr
\hfill\cos\bigg(\frac{\varphi ''_{n_0}\tau}{2}\bigg)\cos\bigg(\varphi '_{n_0}\tau-
\varphi ''_{n_0}\Big(\frac{(qa)^2}{2}-1\Big)
\tau+\frac{(qa)^2}{2}\sin\varphi ''_{n_0}\tau\bigg),
\hfill\llap{(A4)}\cr}
$$
$$
\displaylines{
\overline{v_y}(\tau)=\frac{\hbar q}{m}\exp\bigg(-\Big(qa\sin
\frac{\varphi ''_{n_0}\tau}{2}\Big)^2\bigg)\times\cr
\hfill\cos\bigg(\frac{\varphi ''_{n_0}\tau}{2}\bigg)\sin\bigg(\varphi '_{n_0}\tau-
\varphi ''_{n_0}\Big(\frac{(qa)^2}{2}-1\Big)
\tau+\frac{(qa)^2}{2}\sin\varphi ''_{n_0}\tau\bigg).
\hfill\llap{(A5)}\cr}
$$
As clearly seen from these expressions the velocity oscillations
reduce exponentially with time. The corresponding damping time is
determined by Eq.(29).

\section*{Appendix B}

i) Below we find more detailed expression for wave function (24). In order to do that we perform the
integration over $p$. The corresponding integral has the form
$$
Q_k=\int\limits_{-\infty}^{+\infty}\varphi_p(x)g(p)\phi_k(y-y_c(p))\, dp,
\eqno{(B1)}
$$
where $\varphi_p(x)=(1/\sqrt{2\pi\hbar})\exp(ipx/\hbar)$, where $g(p)$ is determined by
Eq.(22) and $\phi_k(y-y_c(p))$ is the well-known linear harmonic oscillator function.
Putting these expressions into Eq.(B1) and using Gaussian transformation for
Hermite polynomials $H_k(y)$\cite{RG}
$$
\frac{1}{\sqrt{\pi}}\int\limits_{-\infty}^{+\infty}e^{-(x-y)^2}H_k(y)\, dy=(2x)^k,
\eqno{(B2)}
$$
we finally obtain
$$
Q_k=\frac{((y-qa^2-ix)/a)^k}{\sqrt{2^{k+1}k! \pi a^2}}\exp\Big(\frac{2ix(y+qa^2)-x^2-
(y-qa^2)^2}{4a^2}\Big).
\eqno{(B3)}
$$
Substituting the above expression into Eq.(24) and using polar coordinates (32) one
gets Eq.(31).

ii) Let us now calculate the electron wave function $\psi({\bf r},t)=\psi_{cl}({\bf r},t)$ at
the short times $t\ll T_D$ when we can neglect the quadratic term in the expansion
(A2). Then recollecting that for a given wave packet the coefficients $d_n$ and $b_n$
can be approximated by the expressions $d_n\simeq 1$ and
$b_n\simeq\sqrt{n/2}\lambda/a$, we find after summation from Eq.(31)
$$
\displaylines{ \psi_{cl}(\rho,\theta
,t)=\frac{M(\rho,\theta)}{\sqrt{\alpha^2+\beta^2}}S\Big(\theta+\frac{2\pi
t}{T_{cl}}\Big)\exp\Big(-i\Big(\frac{\varepsilon_{n_0}}{\hbar}+\frac{2\pi}{T_{cl}}(1-n_0)\Big)t\Big)\times\cr
\hfill\Big(\alpha,\beta e^{i\frac{2\pi t}{T_{cl}}},\frac{\beta
q\lambda}{2},
-\frac{\alpha\lambda\rho}{2a}e^{-i\theta}\Big)^T,\hfill\llap{(B4)}\cr
}
$$
where
$$
S(u)=\exp\Big(-\frac{qa\rho}{2}(\cos u-i\sin u)\Big)
\eqno{(B5)}
$$
and $M(\rho,\theta)$ is determined by Eq.(33). The solution (B4)
describes the coherent weakly-relativistic wave packet propagating
along the cyclotron orbit without changing its shape. In
particular, at $\alpha=\beta$ the corresponding probability
density has the Gaussian form
$$
|\psi_{cl}(\rho,\theta ,t)|^2=\frac{1+\frac{\lambda^2}{8a^2}(\rho^2+(qa)^2)}{2\pi a^2}
\exp\Big(-\frac{\rho^2+(qa)^2+2\rho qa\cos(\theta+2\pi t/T_{cl})}{2}\Big).
\eqno{(B6)}
$$
It is clearly seen that the center of the wave packet corresponding to the angle
$$
\theta_c=\pi-\frac{2\pi t}{T_c}
\eqno{(B7)}
$$
rotates with cyclotron frequency $\omega_c=2\pi/T_c$ similar to the motion of a
classical electron placed in a magnetic field.

\section*{Appendix C}

Below we show that the wave packet discussed in Sec.IV (Eqs.(47), (48)) at $t_0=T_{cl}/4$
is a coherent superposition of mesoscopic states\cite{{Gea,Berm}}. To do it we consider the
time evolution of the initial state
$$
\psi_{+}({\bf r},0)=\psi_c({\bf r})|+>,
\eqno{(C1)}
$$
where the "orbital" wave function of the coherent state $\psi_c({\bf r})$ is given by Eq.(48)
and the initial Dirac spinor $|+>$ has the form
$$
|+>=d_{n_0}\pmatrix{|\uparrow >\cr 0\cr}+b_{n_0}\pmatrix{0\cr |\downarrow>\cr}
\eqno{(C2)}
$$
with the coefficients $d_{n_0}$ and $b_{n_0}$ are given by Eq.(14). Here as above we assume
that the mean value of $n$ in Eq.(49) is large enough: $\overline{n}=(qa)^2/2=n_0\gg 1$.

The state (C1) evolves as
$$
\displaylines{ \psi_{+}({\bf r},\tau)=\int dp\,
\varphi_p(x)g(p)\Bigg[\sum_{n=1}c_n^{+}\pmatrix{d_n\phi_{n-1}(y-y_c)
|\uparrow>\cr
-b_n\phi_n(y-y_c)|\downarrow>\cr}e^{-i\varphi_n\tau}+\cr
\hfill\sum_{n=0}c_n^{-}\pmatrix{b_n\phi_{n-1}(y-y_c)|\uparrow>\cr
d_n\phi_n(y-y_c)
|\downarrow>\cr}e^{i\varphi_n\tau}\Bigg],\hfill\llap{(C3)}\cr}
$$
where the coefficients $c_n^{\pm}$ are
$$
\displaylines{ c_n^{+}=c_nd_{n_0}d_n-c_{n+1}b_{n_0}b_n,\cr \hfill
c_n^{-}=c_nd_{n_0}b_n+c_{n+1}b_{n_0}b_n.\hfill\llap{(C4)}\cr}
$$
Using the weak dependence on $n$ for the coefficients $d_n$ and $b_n$ and also using
the expression for $c_n$, Eq.(23), we obtain from (C4) at $n\simeq n_0\gg 1$
$$
c_n^{+}=c_n,\quad c_n^{-}=0.
\eqno{(C5)}
$$
Then it is not difficult to show that
$$
\psi_{+}({\bf r},\tau)=\bigg(e^{-i\varphi '_{n_0}\tau}d_{n_0}\pmatrix{|\uparrow>\cr
0\cr}+b_{n_0}\pmatrix{0\cr |\downarrow>\cr}\bigg)\psi_{c+}({\bf r},\tau),
\eqno{(C6)}
$$
where
$$
\psi_{c+}({\bf r},\tau)=\int dp\, \varphi_p(x)g(p)\sum_{n=0}c_{n+1}\phi_n(y-y_c)
e^{-i\varphi_n\tau}.
\eqno{(C7)}
$$
In the same way one can compute the evolution of the initial state
$$
\psi_{-}({\bf r},0)=\psi_c({\bf r})|->,\eqno{(C8)}
$$
$$
|->=b_{n_0}\pmatrix{|\uparrow>\cr o\cr}-d_{n_0}\pmatrix{0\cr |\downarrow>},
\eqno{(C9)}
$$
$$
\psi_{-}({\bf r},\tau)=\bigg(e^{i\varphi '_{n_0}\tau}b_{n_0}\pmatrix{|\uparrow>\cr
0\cr}-d_{n_0}\pmatrix{0\cr |\downarrow>\cr}\bigg)\psi_{c-}({\bf r},\tau),
\eqno{(C10)}
$$
where $\psi_{c-}({\bf r},\tau)$ is determined by Eq.(C7) in which $\exp(-i\varphi_n\tau)$
should be replaced by $\exp(i\varphi_n\tau)$. Note that the initial spin functions are orthogonal:
$<+|->=0$.

Thus, the initial wave packet, Eq.(47), can be represented as a superposition
$$
\psi_c({\bf r})\pmatrix{|\uparrow>\cr 0\cr}=d_{n_0}\psi_{+}({\bf r},0)+
b_{n_0}\psi{-}({\bf r},0),\eqno{(C11)}
$$
so that
$$
\psi({\bf r},\tau)=d_{n_0}\psi_{+}({\bf r},\tau)+b_{n_0}\psi_{-}({\bf r},\tau)
\eqno{(C12)}
$$
describing two sub-packets which at $t<T_{cl}=(2\pi / \varphi '_{n_0})\lambda /c$
rotate counterclockwise $(\psi_{+}({\bf r},\tau))$ and clockwise $(\psi_{-}({\bf r},\tau))$
in the $xy-$plane. At the moment of time $t_0=T_{cl}/4$ as follows from Eqs.(C6) and (C10)
$$
\bigg|<+|->\Big\vert_{t=t_0}\bigg|=\sqrt{\frac{2n_0(\lambda/a)^2}{1+2n_0
(\lambda/a)^2}}\approx 0.98
\eqno{(C13)}
$$
for $n_0=50$ and $\lambda/a=0.5$, and we conclude that the spin parts of wave functions
$\psi_{\pm}({\bf r},t)$ differ slightly. That means that "orbital" part of full wave function
(C12) at that time is a linear superposition of the mesoscopic states $\psi_{c+}({\bf r},t_0)$
and $\psi_{c-}({\bf r},t_0)$ (the so-called "Schr\"odinger cat"). The corresponding probability
density $|\psi({\bf r},t_0=T_{cl}/4)|^2$ is shown in Fig.7b. The generation of coherent
superposition of mesoscopic states for the Dirac particle placed in the magnetic field was predicted
in Ref.\cite{Berm}.

\end{document}